\documentclass[nonacm]{acmart}

\AtBeginDocument{%
  }

\usepackage{colortbl}
\usepackage{multirow}
\usepackage{soul}
\usepackage{booktabs}
\usepackage{array}
\usepackage{graphicx}
\usepackage{svg}

\usepackage[normalem]{ulem}
\usepackage{xcolor}

\begin{document}

\title[``The Explanation Makes Sense'']{``The Explanation Makes Sense'': An Empirical Study of LLM News Classification and Its Influence on Human Judgment}

\newcommand{\udaffiliation}{
  \affiliation{%
    \institution{University of Delaware}
    \city{Newark}
    \state{DE}
    \country{USA}
  }
}

\author{Qile Wang}
\orcid{0000-0003-0308-6033}
\email{kylewang@udel.edu}
\udaffiliation

\author{Prerana Khatiwada}
\orcid{0009-0008-6965-9504}
\email{preranak@udel.edu}
\udaffiliation

\author{Avinash Chouhan}
\orcid{0009-0007-6939-4668}
\email{avinashc2104@gmail.com}
\udaffiliation

\author{Ashrey Mahesh}
\email{mahesha@udel.edu}
\orcid{0009-0007-5652-3694}
\udaffiliation

\author{Joy Mwaria}
\email{jmmwaria@gmail.com}
\orcid{0009-0008-8343-3595}
\udaffiliation

\author{Duy Duc Tran}
\email{dustintr@udel.edu}
\orcid{0009-0008-6983-6312}
\udaffiliation

\author{Kenneth E. Barner}
\orcid{0000-0002-0936-7840}
\email{barner@udel.edu}
\udaffiliation

\author{Matthew Louis Mauriello}
\orcid{0000-0001-5359-6520}
\email{mlm@udel.edu}
\udaffiliation

\renewcommand{\shortauthors}{Wang et al.}

\begin{abstract}
The spread of media bias is a significant concern as political discourse shapes beliefs and opinions. Political news bias classification is inherently subjective, depending on reader interpretation of tone, framing, and language. While Large language models (LLMs) can assist at scale, their influence on human judgment remains poorly understood. To advance understanding of how AI-generated explanations influence subjective judgment, we investigate how LLMs classify U.S. news by political ideology and examine how their explanations shape human decision-making. We first compared GPT models with prompt engineering to state-of-the-art supervised machine learning on a 34k public dataset. We then collected 17k news articles and tested GPT-4 predictions with Brief and Detailed explanations. In a between-subjects study (N=124), we examined how LLM-generated explanations influence human judgment and confidence. Results show that AI assistance significantly increases confidence ($p<.001$), regardless of whether GPT predictions are reliable. Detailed explanations were associated with about three times higher odds of decision switching than Brief explanations, suggesting that explanations may support interpretive reasoning while also increasing confidence and agreement with AI, even when they diverge from publisher-level labels.  Through thematic analysis, we identify the benefits and risks of AI explanations in subjective decision contexts, offer recommendations for responsible AI design, and release our dataset to support future research.
\end{abstract}



\maketitle

\section{Introduction} 
\label{Intro}
\begin{figure}[!ht]
    \centering
    \includegraphics[width=0.9\textwidth]{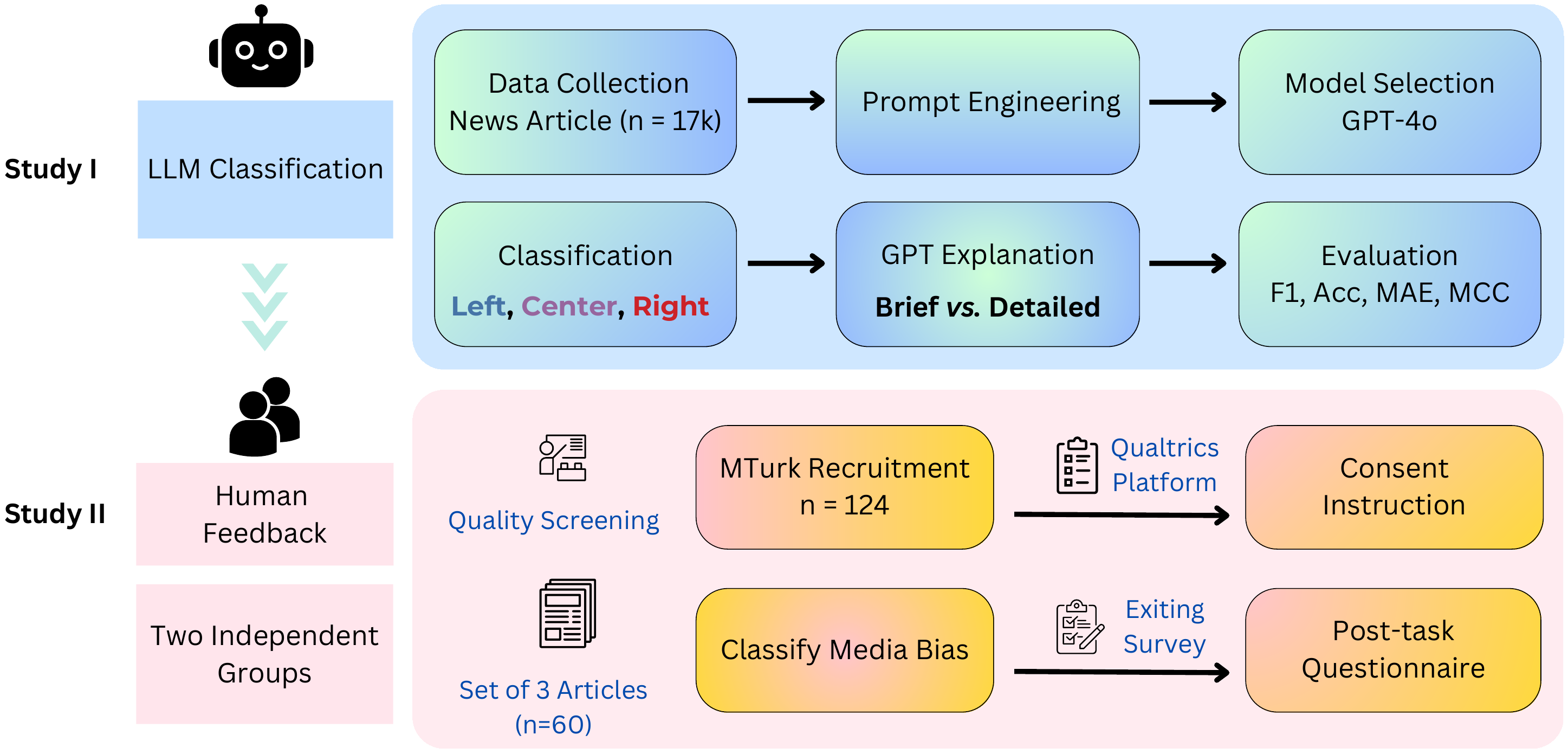}
    \caption{Study Design Overview - This diagram illustrates the key components and workflow of our study, including GPT prediction experiments on news articles and human study procedures}
    \label{fig:overall_flow}
\end{figure}

News media bias is particularly concerning in today's environment, where people often receive news from a limited range of sources that may confirm their pre-existing beliefs \citep{xiang2007news,flaxman2016filter}. This creates echo chambers, where people see fewer opposing views, making public discussions more polarized \citep{dahlgren2020media, bruns2017echo}. As the information landscape continues to evolve, the need for accurate and efficient methods to classify news content becomes increasingly critical \citep{kaur2016news}. 
This is especially true in political discourse, where the choice of words holds substantial power to influence public opinions and perspectives \citep{ecker2022psychological, bryanov2021determinants}.
With effective implementations, artificial intelligence (AI) has the potential to assist in ideology prediction and stance detection \citep{lin2024indivec}. 
Large language models (LLMs) such as GPT-4 have the potential to improve this process by analyzing and generating text that resembles human writing \citep{liu2023summary, orru2023human, khatiwada2026ai}. Recent developments in LLMs have demonstrated their potential in various applications, from content generation, text classification~\citep{chae2025large, wang2025leveraging, wang2026wisdom, wang2026echoes}, sentiment analysis~\citep{kumawat2021sentiment, zhang2023enhancing}, qualitative analysis~\citep{meng2024exploring, wang2025lata} to visualization design~\citep{kim2025good}.
Nevertheless, LLMs tend to perform less effectively on subjective tasks compared to more objective tasks, as the former involves interpretation and emotional nuances, making them more complex \citep{wang2024reasoning}. 
To help reduce media bias and promote more informed and constructive public engagement, we examine how AI-generated labels and explanations influence human judgment, driven by growing interest in the prospective risks and advantages of AI assistance \citep{jain2023effective, dellermann2021future, reverberi2022experimental}.

In addition, many researchers focus on combining the strengths of human-AI collaboration to achieve optimized joint decision outcomes \citep{10.24251/hicss.2022.029, jiang2023beyond}.
However, the effectiveness of LLMs in replicating human-derived labels, particularly in the nuanced task of news content classification, remains an area of active exploration \citep{abburi2023generative, fatemi2023evaluating, guven2024performance}.
Trust in AI continues to be a significant concern \citep{omrani2022trust}. Often, AI-assisted decision-making is influenced by factors such as confidence levels and the provision of explanations for AI recommendations \citep{zhang2020effect}.
Past work \citep{zhang2020effect, wang2023watch, wang2024human, ma2024beyond} has focused on AI-assisted decision-making in tasks that are less subjective. These studies show mixed results on how LLM explanations or confidence scores affect performance across different tasks and domains. Motivated by this, we extend this work to more subjective tasks with a focus on media bias. Specifically, we refer to this decision-making task as making a judgment by categorizing a news article based on its political ideology, a terminology widely used in other similar studies \citep{tuncer2022exploring, wang2021explanations, jain2023effective, xu2024good}. This shift in focus is not only more intricate but also offers valuable insights into how AI can influence human judgment in nuanced situations. Our goal is to explore these dynamics within a controlled and moderated environment, improving our understanding of how a single exposure to AI-generated labels and explanations shapes judgment in a static annotation task, rather than sustained or iterative human-AI collaboration.

Since political news bias classification is inherently subjective, we investigate how LLM-generated explanations influence subjective judgment and confidence in ambiguous domains where accuracy may be difficult to define. Therefore, we focus on three exploratory research questions: First, \textit{(RQ1) what is the performance of LLMs in multi-classification tasks, particularly in identifying media bias?} Second, \textit{(RQ2) how do LLM-generated explanations impact human judgment and self-reported confidence level in a subjective political interpretation task?} \textit{ And, third (RQ3), what are the underlying reasons influencing how users interpret and respond to AI explanations in identifying political ideology?} To address these questions, we first applied GPT to classify a historical news dataset \citep{baly2020we} and compared its performance with state-of-the-art supervised machine learning models. 
We then collected 17,166 news articles covering recent events up to July 2024 and tested GPT-4 predictions with two types of generated explanations. 
Following the analysis, we conducted a human study with 124 participants, each instructed to classify news articles and report their confidence level for each decision both before and after receiving AI-generated information.
Figure \ref{fig:overall_flow} illustrates the overall study design. With prompt engineering, GPT achieves moderate performance on our recent news dataset. We found that GPT explanations significantly increase human confidence levels ($p<.001$), regardless of alignment with publisher-level labels, indicating that explanation-induced influence operates independently of label agreement. The Detailed AI label--explanation condition was associated with stronger persuasive effects than the Brief condition, with participants about three times more likely to change their decisions ($p<.01$). In addition, we explore the reasoning behind human decisions that align with or diverge from AI predictions using a systematic thematic analysis. We investigate user perception, which refers to how individuals interpret and evaluate AI information based on their experiences, understanding, and attitudes toward AI capabilities and reliability.

The main contributions of this paper are summarized as threefold. First, we provide the first systematic investigation of how different LLM explanation formats (Brief versus Detailed) shape user confidence, decision-making, and reliance in a subjective political annotation task. Our findings demonstrate a critical dual effect in which Detailed explanations may provide interpretive support while also increasing reported confidence, even when they diverge from publisher-level labels. This reveals that the persuasiveness of explanations can operate independently of alignment with publisher-level reference labels. Second, we provide empirical evidence that explanation-induced influence in subjective domains presents a fundamental design tradeoff: while explanations support reflective judgment, they simultaneously increase reliance and confidence in ways that may not be justified. Finally, we identify design implications and propose principles for AI systems in subjective domains that prioritize transparency about ambiguity, support for reflection over reliance, and calibrated confidence in user judgment.



\section{Related Work} 
\label{related_work}
We review existing research on media bias, LLMs in classification, crowdsourced data collection, and the impact of human-AI collaboration on user judgment, highlighting how these elements contribute to improving the efficacy and reliability of AI systems in mitigating news bias.

\subsection{News Credibility and Media Bias}
News credibility is a major concern because it affects how much people trust the media and how they make decisions \citep{mehrabi2009news}. The rapid propagation of misinformation, coupled with the challenge of distinguishing facts from biased narratives, exacerbates this issue \citep{vosoughi2018spread, morstatter2018identifying}. Individuals' perceptions of news credibility typically rely on their personal beliefs and the manner in which the information is conveyed \citep{metzger2010social, flanagin2000perceptions}.
Furthermore, variations in linguistic structure, rhetorical emphasis, and thematic organization can lead audiences to interpret the same news article in meaningfully different ways \citep{pan1993framing}. These dynamics underscore the need to examine news credibility not only as a factual issue, but also as a subjective and interpretive process shaped by cognitive, ideological, and contextual factors.

Media bias adds to these concerns and describes how news may favor certain interests, such as the government \citep{gehlbach2014government}. Prior work highlights several forms of bias, such as visibility bias, tonality bias, and agenda bias. These forms shape which voices gain attention and which issues receive prominent coverage \citep{eberl2017one,puglisi2015empirical}. Propaganda analysis relates to these patterns because it examines how media can guide public perception through emotional or psychological techniques such as loaded language \citep{weston2018rulebook}, flag waving \citep{hobbs2014teaching}, straw man arguments \citep{bentham1996straw}, and red herrings \citep{weston2018rulebook, teninbaum2009reductio}. This line of work helps explain why media bias is complex and often hard to measure.
Annotators often disagree on whether an article is biased, its direction, and its intensity, differences driven by ideological priors, tone interpretations, and the inherent ambiguity of political language \citep{liu2010sentiment}. Such disagreement is not mere noise but reflects meaningful interpretive variation across readers \citep{sandri2023don}. Similar patterns appear in argument mining and political debate analysis, where crowdsourced labels show high variability and low agreement \citep{mestre2022benchmark}. As a result, assessing media bias relies heavily on subjective judgment rather than definitive ground truth, making it more complex than typical fact-based classification tasks.

These challenges in identifying and evaluating bias have important downstream consequences. Media bias is closely intertwined with political polarization, as biased coverage can deepen divides in public opinion \citep{hart2020politicization}.
\citet{tucker2018social} demonstrate that rising polarization fueled in part by biased or selective media environments has contributed to increased political misinformation, leading to misconceptions and conspiracy theories. Biased reporting not only shapes public interpretations of events but also influences individual behavior and collective decision-making \citep{hamborg2019automated, krieger2022domain}. 
Such bias in news can significantly and systematically alter reality, affecting the way news readers perceive information \citep{groeling2013media}. 
Raising awareness of media bias is therefore essential. Prior efforts have attempted to support news readers by communicating or visualizing bias, such as through visual interfaces that highlight biased language in articles \citep{spinde2022we} or analytic tools like Newsalyze \citep{hamborg2021newsalyze}.
Motivated by these challenges, we aim to leverage LLMs for a more generalized approach and investigate their effectiveness in collaboration with human labelers to promote awareness of media bias and interpretation in high-ambiguity political news contexts.

\subsection{Classification Tasks - MTurk and LLMs}

Crowdsourcing platforms such as Amazon Mechanical Turk (MTurk) have become central to large-scale annotation and classification tasks across domains \citep{rashtchian2010collecting, kutlu2020annotator, kasthuriarachchy2021cost, soprano2024longitudinal}. Prior work highlights MTurk’s utility for building training sets, conducting content analysis, image classification, and running online surveys \citep{leeper2016crowdsourced, egbert2015developing}. Clear labeling instructions are essential to ensure crowdworker quality \citep{radsch2023labelling}. These advantages make MTurk a scalable, cost-effective choice for gathering human judgments \citep{wang2013perspectives}, motivating our use of the platform to study human interpretations of political news.

While MTurk is widely used, recent work shows that subjective annotation tasks (such as media bias, political stance, and contextual online harms), particularly in political domains, naturally produce systematic disagreement driven by differences in values, experiences, and interpretations rather than participants' error \citep{davani2022dealing}. 
%
%
Recent work instead treats disagreement as a meaningful signal, developing models that preserve annotator perspectives and model uncertainty \citep{mokhberian2024capturing}. Yet little is known about how such interpretive variation unfolds in human–AI collaboration. Addressing this gap, we use MTurk not only for data collection but also to observe how individuals interpret political news when assisted by LLM explanations, allowing us to analyze how AI reasoning shapes human judgment and confidence.

Previous unsupervised modeling and extraction methods analyzed selected topics in news media (abortion, immigration, and gun control) using entity-level polarization annotation \cite{paschalides2025framework}. However, it was a very limited case study, which may restrict the generalizability of the findings.
Large Language Models (LLMs), meanwhile, have shown strong capabilities across classification and prediction tasks in news domains without topic constraints \citep{fatemi2023evaluating, alonso2023framing}, supporting applications such as categorization, retrieval, summarization, and recommendation \citep{chae2025large, zhang2024benchmarking}.
Although LLMs have shown strong performance in objective downstream tasks such as income prediction from user profiles \citep{zhang2020effect}, poisonous mushroom identification \citep{wang2023watch}, natural language inference-based annotation \citep{wang2024human}, and topical news classification \citep{ma2024beyond}, far less is known about how humans interpret AI-generated information in high-ambiguity settings. We contribute a perspective that extends beyond objective labeling tasks and offers new insight into how humans rely on and respond to AI explanations in complex political contexts.

\subsection{Human-AI Collaboration}
Here, we provide an overview of how human-AI interaction dynamics are studied, including the benefits and challenges of collaboration between humans and AI systems.

\subsubsection{Social Influence, Collaborative Dynamics, and Human-in-the-Loop AI}
Human-in-the-loop (HITL) systems have become central to collaborative annotation, dataset curation, and design tasks, where combining human judgment with AI assistance improves efficiency and data quality \citep{subramonyam2022solving, russakovsky2015best, yuan2021synthbio}. This collaborative approach uses human cognitive abilities and AI computational power, making it possible to handle complex tasks more efficiently and leading to optimized decision outcomes \citep{zhang2020effect}. For instance, \citet{kim2024meganno+} introduced MEGAnno+, which integrates LLM suggestions with human verification, while \citet{zhang2024tripartite} demonstrated that combining human judgment, LLMs, and neural networks enhances misinformation classification. Similar benefits appear in content moderation, where AI can aid humans by surfacing likely problematic content, while humans provide the necessary contextual interpretation \citep{lai2022human}. HITL approaches have also been shown to influence user perceptions and decision-making in consumer contexts \citep{yue2023impact}.

\begin{table}[!h]
    \small
    \caption{Prompt Design and Development}
    \label{table:prompt_design}
    \centering
    \begin{tabular}{cm{14cm}}
        \toprule
        \textbf{\#} & \textbf{Prompt Structure} \\ 
        \midrule
        A & Your task is to ACCURATELY CLASSIFY the news into one of the following categories: Left, Center, or Right. \textit{\textbf{Left} represents strong alignment with liberal, progressive, or left-wing thought and/or policy agendas.} \textit{\textbf{Center} represents publications that do not show significant political bias or display a balance of articles with both left and right perspectives.} \textit{\textbf{Right} represents strong alignment with conservative, traditional, or right-wing thought and/or policy agendas.} Only return Left, Center, or Right. \{News Content\} \\ 
        \midrule
        B & You are a political expert and you will help to accurately classify the news into one of the following political bias: \textit{\textbf{Left}: News that strongly aligns with liberal, progressive, or left-wing thought and/or policy agendas.} \textit{\textbf{Center}: News that maintains a neutral stance or balances perspectives from both left and right viewpoints.} \textit{\textbf{Right}: News that strongly aligns with conservative, traditional, or right-wing thought and/or policy agendas.} Output Instruction: Return only one of the following labels: Left, Center, or Right. \{News Content\} \\ 
        \midrule
        C & Your role is to accurately classify the news content into one of the following categories based on its alignment with specific thought and policy agendas, with particular attention to the tone and overall favorability toward these agendas. \textit{\textbf{Left Bias Indicators}: Favor for generous government services (e.g., food stamps, social security, Medicare), Advocacy for rejecting social and economic inequality ...}[12 more indicators omitted]... \textit{\textbf{Center Bias Indicators}: Either do not show much political bias or display a balance with left and right perspectives. Support for moderate government intervention and balanced budgets. Media organizations: BBC News,} ...[9 more media source]...\textit{ \textbf{Right Bias Indicators}: Emphasis on freedom of speech and traditional family values. Advocacy for decreasing taxes and preserving gun rights} ...[12 more indicators]... Instructions: 1. Carefully review the content of the article. 2. Determine the general political bias based on the overall tone and favorability toward left or right thought and policy agendas 3. Consider the media organization if mentioned 4. IMPORTANT: Only return one of these three words: Left, Center, or Right.  \{News Content\}\\ 
        \midrule
        D & Your task is to ACCURATELY CLASSIFY the news into one of the following categories: Left, Center, or Right. \textit{\textbf{Left} represents strong alignment with liberal, progressive, or left-wing thought and/or policy agendas. Media organizations that are usually Left-biased: AlterNet, The Atlantic}...[30 more omitted]... \textit{\textbf{Center} represents publications that do not show significant political bias or display a balance of articles with both left and right perspectives. Media organizations that are usually center: BBC News, Forbes} ...[8 more]... \textit{\textbf{Right} represents strong alignment with conservative, traditional, or right-wing thought and/or policy agendas. Media organizations that are usually right: The American Conservative, CBN}...[22 more]... Only return Left, Center, or Right. \{News Content\} \\ 
        \midrule
        E1  & Based on the bias (or lack thereof) in the article, does this text lean ``left'' or``right'' or ``center''? Explain specific elements that indicate bias and justify your choice. Article Text for analysis: \{News Content\} Format of Output: (left, right or center), (explanation) \\
        \cline{2-2}
        E2 & You previously labeled this article as \{GPT Bias Label\} and provided the following explanation: \{GPT Explanation\} Although your answer may be right, BE CRITICAL OF THIS ANALYSIS. Bias is NOT about the selection of events covered but RATHER about the writing itself. If you notice the following, the source may be biased: Heavily opinionated or one-sided. Relies on unsupported or unsubstantiated claims. Presents highly selected facts that lean toward a certain outcome...[6 more guides omitted]... Analyze this Article Text Again: \{News Content\}. Do you still agree with the initial bias label (remember: be critical of yourself)? If not, provide a revised label and explanation.\\
        \bottomrule
    \end{tabular}
\end{table}

Beyond task performance, recent work in human--AI collaboration foregrounds the social and perceptual dynamics that arise when people work with AI systems. In cooperative game settings, a study \cite{ashktorab2020human} shows that social perceptions such as likeability, intelligence, and creativity differ depending on whether participants believe they are collaborating with an AI or another human, even when underlying behavior is held constant. This aligns with social influence research demonstrating that individuals adapt their judgments after exposure to others' opinions and confidence levels, with both “expert effects” and “majority effects” shaping collective outcomes \citep{moussaid2013social}. In collaborative learning contexts, peer interactions and socially shared regulation of learning (SSRL) further highlight how group processes regulate individual motivation and engagement over time \citep{jarvenoja2025investigating}, reinforcing that group members' beliefs, goals, and regulatory moves can systematically steer each other's cognitive and motivational states. Complementary work on human--AI collaboration in qualitative analysis shows that scholars are willing to incorporate AI as long as it supports, rather than replaces, human interpretive judgment and control over analytic decisions \citep{feuston2021putting}. Building on these insights, our study aims to explore human-in-the-loop interactions with AI for detecting and mitigating media bias through a controlled, pilot testing approach.

\subsubsection{AI-Assisted Decision-Making}
Despite rapid advances in AI, significant challenges remain due to misunderstandings of AI capabilities, social-context limitations, and technical constraints \citep{hagendorff202015}. These challenges can create disparities in decision-making and reasoning, leading to inconsistencies and trust issues \citep{chang2023examining, xu2024good}. User perceptions of AI reliability strongly influence their willingness to adopt AI systems \citep{rabi}. Factors such as credibility, risk, ease of use, perceived integrity, and fairness shape trust, which in turn underpins broader attitudes toward AI technologies \citep{bach2024systematic, yang, choung2025fairness}. Human-like or socially responsive AI can enhance reliance by offering interpretable, conversational interactions \citep{bostrom, tuncer2022exploring}, whereas perceived errors, reduced control, or high risk often lead to disengagement \citep{tang}. Trust is further modulated by confidence cues, explanation quality, workload, satisfaction, and performance expectancy \citep{zhang2020effect, choudhury}, with transparency especially critical in high-stakes contexts \citep{ashoori2019ai}.

\begin{figure}[!htb]
    \centering
    \includegraphics[width=0.95\textwidth]{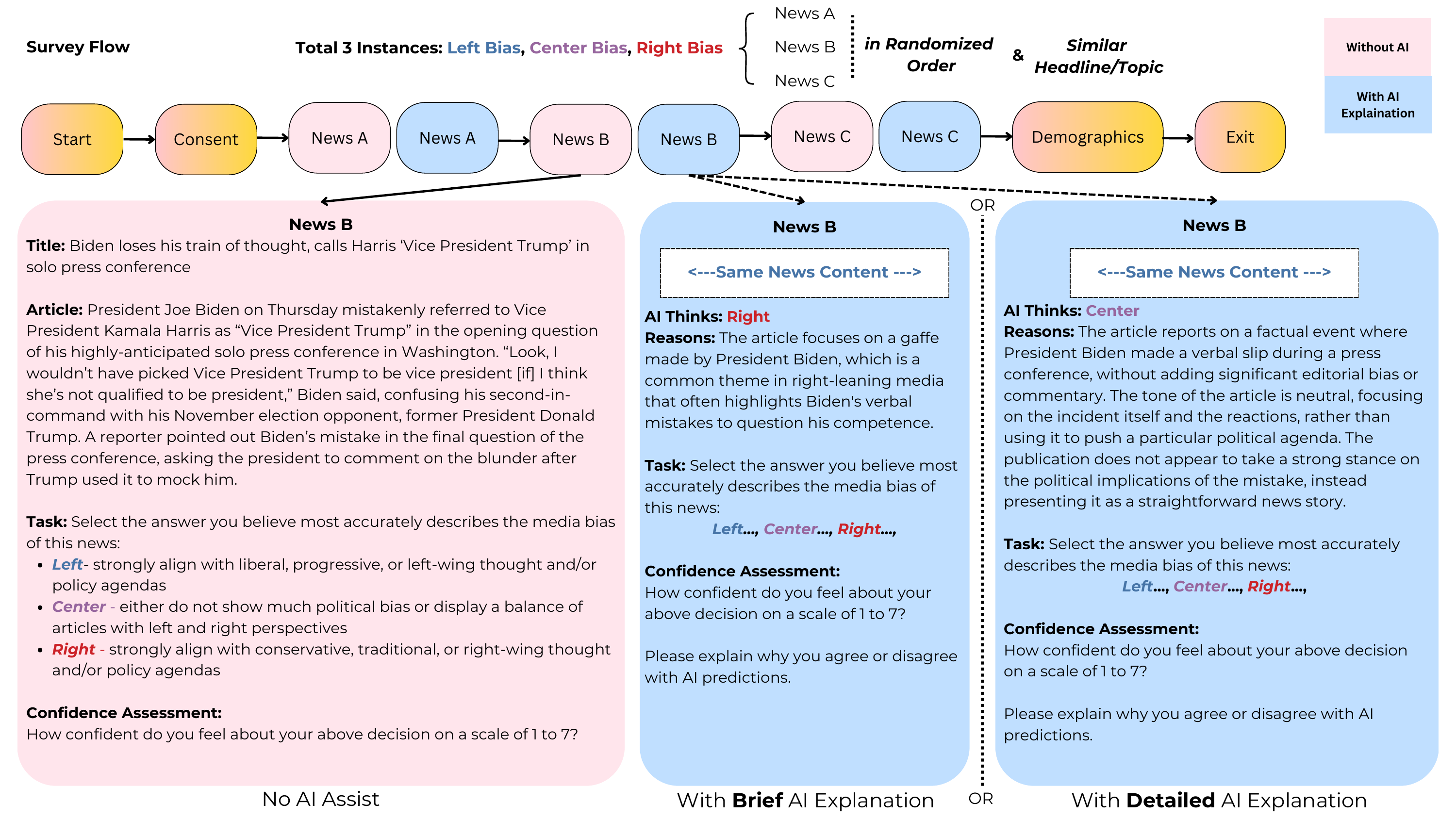}
    \caption{Survey Flow and User Interface for Classifying Media Bias Tasks. The sample news article shown in the figure is categorized as \textbf{Center} based on the data collection process. Each participant reads three articles from similar stories, one left, one center, and one right, shown in random order. Because the Brief and Detailed outputs were generated in separate GPT requests, the same article can receive different GPT labels in the two conditions, as in this example (Right with the Brief explanation, Center with the Detailed explanation); the longer reasoning required for the Detailed explanation may lead the model to a different classification (Section \ref{sub:Design}).}
    \label{fig:interface}
\end{figure}

LLMs generate explanations that can enhance subjective trust when integrity-focused \citep{integrity}, yet their effectiveness varies with domain expertise, task type, and explanation format \citep{wang2021explanations, wang2024human}. Studies have found that the explanations do not always increase accuracy and may even induce over-reliance on incorrect suggestions \citep{liu2018towards, bansal2021does}. Human verification of LLM outputs can improve annotation quality; however, outcomes remain inconsistent, highlighting the need for further research. Rather than focusing on objective accuracy, we examine how LLM-generated explanations shape human confidence, reliance, and interpretive judgment. By framing explanations within theories of social influence and collaborative regulation, we treat LLM outputs as an additional “voice” in small interpretive groups, where rationale and implied confidence function like expert and majority cues \citep{moussaid2013social, jarvenoja2025investigating}. This approach highlights how explanation framing can steer opinion formation, revealing opportunities and risks that objective labeling studies may overlook.

\subsubsection{Bias in AI}
Although AI has great potential to assist human experts, users often remain hesitant due to concerns about accuracy, bias, reliability, and the potential impact on critical outcomes \citep{zhang2020effect}. Skepticism also arises from fears that AI may not fully understand context, handle subtleties, or may perpetuate inequalities and harmful stereotypes \citep{ferrara2024fairness}. Conversely, human annotators can introduce biases that AI might avoid \citep{gemalmaz2021accounting}. Therefore, building trust in AI requires transparency, accountability, and ethical practices throughout development and deployment.

Beyond user perceptions of AI reliability, the internal reliability of LLMs themselves is critical, particularly the extent to which they encode and propagate biases from large-scale training corpora. Foundational work shows that LLMs inevitably absorb societal and ideological biases, which can distort outputs and amplify misinformation risks \citep{bender2021dangers, weidinger2021ethical}. Empirical studies demonstrate that model responses reflect latent priors and answer preferences \citep{zhao2021calibrate}, and that LLM-generated political opinions do not reliably align with real demographic groups \citep{santurkar2023whose}. In news and political contexts, such biases can shape classification, framing, and generated explanations, including reproducing media frames \citep{horych2025promises}, skewing coverage by gender or race \citep{fang2024bias}, or exhibiting ideological lean in political bias ratings \citep{hernandes2024llms}. These findings suggest that user trust and decision-making may be influenced not only by how explanations are presented, but also by the underlying biases embedded in the model itself.

Building on these prior works, we explore how LLM, together with the framing of generated explanations, shape user confidence and decision-making in subjective news-interpretation tasks. While much existing research focuses on objective labeling or annotation performance, we examine scenarios requiring contextual understanding and personal judgment. Motivated by mixed evidence on the effectiveness of GPT-generated explanations, we introduce two formats, Brief and Detailed, to evaluate how explanation design influences user reliance and perceived reliability. This approach enables us to investigate how the presentation of explanations interacts with underlying model behavior to influence interpretive decisions, providing new insights into supporting subjective reasoning tasks while accounting for both presentation and model bias effects.


\section{Method}
\label{methods}

In Study I, we discuss the GPT experiments and various evaluation metrics, focusing on performance in media bias classification and explanations. In Study II, we detail participant recruitment, demographics, and the human study design. Figure \ref{fig:overall_flow} illustrates our overall study design.

\subsection{Study I - GPT Classification Performance}
First, we describe our GPT classification process, including dataset collection, prompt engineering, and experiment execution, before delving into the human study.

\subsubsection{Dataset}
We used two datasets for our GPT experiments, both evaluated using the left-right political spectrum 
analyzed from AllSides \footnote{\url{https://www.allsides.com}}, a platform with the goal of raising awareness of media bias and providing diverse perspectives on similar news stories from different sources.
We first conducted initial experiments using the existing dataset from \citet{baly2020we}, which contains 34,737 full-text news articles originally sourced from AllSides and were manually annotated and verified with political leanings (Left, Center, Right) at the article level; they found that approximately 97\% of article bias labels matched the bias of the publisher. Each article has an average of 968 words (SD = 690).
The articles were published between July 2012 and June 2020. We first use this dataset for GPT model selection and prompt engineering in order to evaluate GPT’s potential compared with state-of-the-art supervised methods.

For the purpose of conducting human studies, we create a new dataset that has more recent news. This is because people are generally more familiar with current events, particularly those from 2024. Therefore, we collected a total of 17,166 articles covering 5,722 round-up news headlines with 64 unique categorical topics sourced from AllSides using a public scraping tool\footnote{\url{https://github.com/csinva/news-title-bias/tree/master}}. Modifications to the script were necessary due to changes in the AllSides website’s HTML structure. Each news round-up includes three similar stories: one from left or left-leaning media, one from center media, and one from right or right-leaning media. Our collected dataset includes the news titles, news categorical topics, publication dates, publisher sources, headline roundup titles, political leanings (on a 5-point scale), URLs, and snippets of the news stories. 
In addition, our dataset contains news covering July 2012 to July 2024, with the majority of topics on "Politics," "Elections," "Economy and Jobs," "World," "Coronavirus," etc.  

Each article inherits the leaning that AllSides assigns to its publisher. This convention is simple, transparent, and reproducible, so we use it as the reference for all analyses and for future studies that use the same source. We note that political ideology classification is inherently subjective, and publisher-level labels serve as reference conventions rather than objective ground truth. Different readers may reasonably interpret the same article's political framing differently, which is why we use the term ``publisher-level reference labels'' throughout this paper and refer to alignment or divergence from these labels rather than using terms implying objective correctness.

In today's digital landscape, news consumption is often characterized by brief and rapid interactions, where readers frequently skim headlines, prioritize key content, or engage with short summaries (i.e., hyper-reading behavior) rather than reading full-length investigative reports \citep{duggan2009text, waage2018hyper}. To reflect this real-world behavior, our study was designed to align with how individuals typically encounter and interpret news content. This approach reduces cognitive load and enhances participant engagement, reflecting real-world news consumption \citep{hollender2010integrating}.
Each news snippet in our collected dataset consists of a summary or the first paragraph of an article, averaging 77 words. Records with fewer than 10 words were excluded due to factors such as video-only content, inaccessible media, or restricted access.

\begin{figure}[!htb]  
    \centering
    \includegraphics[width=0.6\textwidth]{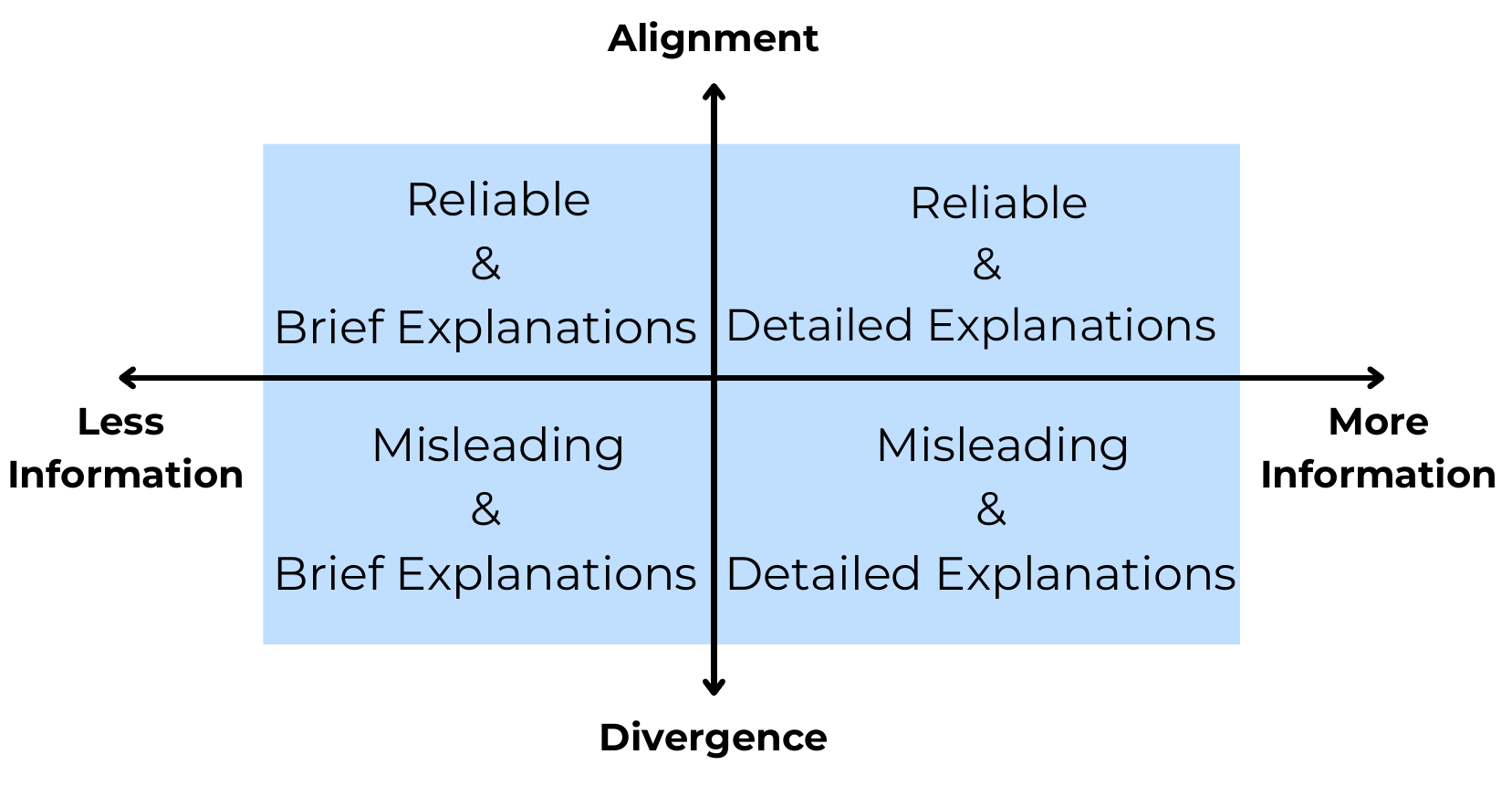}  %
    \caption{Experimental Conditions Defined by AI Alignment and Information Level. The vertical axis shows the alignment between LLM predictions and publisher-level reference labels, and the horizontal axis represents explanation depth. In the figure, `Reliable' denotes alignment with and `Misleading' denotes divergence from the publisher-level reference labels.}
    \label{fig:quad}
\end{figure}

\subsubsection{Experiment Design}
\label{prompt_deisgn}
Our experiment focused on testing various GPT models offered by the OpenAI\footnote{\url{https://openai.com}} API to determine the performance of bias classification tasks. We began with GPT-3.5 (``gpt-3.5-turbo-0125'') as a standalone model. Subsequently, we incorporated GPT-4o (``gpt-4o-2024-05-13''), the most advanced model at the time, and GPT-4o (``gpt-4o-2024-08-06''), known for its support of structured outputs as an additional feature. Furthermore, we experimented with a smaller but cost-effective model, GPT-4o mini (``gpt-4o-mini''). 
The \textit{Temperature} setting controls the randomness in GPT output. A low value gives predictable text. Lower \textit{Temperatures} often lead to more predictable and repetitive outputs, while higher \textit{Temperatures} might result in less predictable outputs but introduce greater diversity. We tested different values of in the segment starting from 0, since this is the most deterministic setting.

In initial experiments, we instructed GPT to classify the political leaning of news articles using only the news content. In other experiments, we included an additional news title as input. The media source name was never provided. This approach reduces the risk of GPT learning patterns specific to sources, which could obscure its ability to capture the underlying political ideology from the news content itself. Given that GPT has been trained on large datasets that likely include media source biases, withholding the source information helps mitigate reliance on such biases. Similarly, we did not provide the media source to participants for the next part of a human study to prevent prior knowledge of source bias from influencing their judgments and to ensure they focus more on the content of the articles.

\textbf{Prompt Engineering.}
To construct prompts for GPT models, we conducted iterative prompt engineering inspired by \citet{bsharat2023principled}. 
Table \ref{table:prompt_design} shows the three main types of prompts used in our experiments. Bold and italic formatting in the table is only for visual emphasis and does not appear in the actual prompts. First, Prompts A and B are short prompts that include only label definitions. Secondly, Prompt C gives more guidance with detailed bias indicators, including media sources linked to each political ideology class. Prompt D utilizes only the source media bias label, in addition to Prompt A. These descriptions come from AllSides\footnote{\url{https://www.allsides.com/}}. Thirdly, the LLM generates a label and an explanation using Prompt E1, and subsequently the LLM reviews the previous answer using Prompt E2. This self-critique design helps reduce hallucination and allows a check on the model’s initial output.

\textbf{Evaluation.}
For evaluation, we assessed performance using several metrics. Precision, Recall, F1 Score, and Accuracy were measured, with values closer to 1 indicating better performance. We also computed the Mean Absolute Error (MAE) by assigning numerical values to the classes—left as 0, center as 1, and right as 2. A smaller MAE, closer to 0, indicates higher accuracy. We also used the Matthews Correlation Coefficient (MCC) to indicate model quality, where the coefficient ranges from -1 to 1, with 1 representing the optimal classification model.
Additionally, we measured the quality of the explanations using the TextDescriptives package \citep{hansen2023textdescriptives} and focused on readability and perplexity metrics. 

\subsection{Study II - Crowdsourcing MTurk Study: Human Feedback}


\subsubsection{Recruitment and Data Collection}
To understand decision-making and the impact of AI assistance on bias classification, we recruited 124 participants (two groups of 62) via MTurk in late August 2024. This study received approval from the Institutional Review Board of the affiliated university 
(Protocol \#2037012-3).
Participants signed a digital consent form, completed classification tasks, and then completed a demographic survey. The entire study was conducted using a between-subjects design and streamlined through Qualtrics, as outlined in Figure \ref{fig:interface}.
To ensure English proficiency and data quality, participants were required to be based in the U.S. and have a HIT approval rate of at least 90\% to qualify for the task.
Sample size requirements were determined using G*Power \cite{faul2007g} for a mixed-design ANOVA (within--between interaction). 
An \textit{a priori} power analysis ($f = 0.25$, $\alpha = 0.05$, $1-\beta = 0.95$) indicated a required sample size of $N = 54$ participants for a mixed $2 \times 2$ design with explanation type (Brief vs Detailed) as a between-subject factor and alignment of LLM predictions with publisher-level reference labels (aligned vs divergent) as a within-subject factor (Figure~\ref{fig:quad}).
To improve the reliability and accuracy of effect estimation, we recruited 124 participants, which reduces the risk of Type I and Type II errors.
We collected demographic information anonymously, including political leaning, age, gender, race, education, etc. Additionally, participants were asked whether they use AI for annotation tasks. We also included an open-ended question to gather participants' experiences with AI annotation tools and systems.

\subsubsection{Assessment and Refinement of Response Quality}
\label{section:quality}
To ensure high data quality, several measures were implemented throughout the study. First, we incorporated a time delay mechanism for the appearance of the ``Next'' button in our survey in each task to encourage participants to read the news article and information provided, reducing the likelihood of content being skipped. Additionally, bot detection features in Qualtrics were used to filter out non-human responses to ensure that only genuine participants contributed to the data. We reduced the number of open-ended questions to just one per task to avoid tiring out participants.
We conducted a thorough quality check of qualitative data to ensure that the responses were meaningful and relevant. None of the responses in the study needed to be filtered out.
We launched our human study at an optimal time and day (during the morning) to reach more participants with better engagement. Moreover, open-ended questions required a minimum of 50 characters to ensure the quality of our thematic analysis. Demographic questions were positioned at the end of the survey to prevent fatigue, ensuring participants were most attentive when answering these critical questions. Additionally, we cross-checked and ensured that all 124 participants were unique and had only completed the task once.

\subsubsection{Design and Execution of Human Classification Task}
\label{sub:Design}
Our study variables and conditions are illustrated in Figure \ref{fig:quad}, where we focus on two conditions: (1) alignment of GPT predictions with publisher-level reference labels, and (2) AI assistance with two types of explanations (Brief and Detailed). 
The term explanation here means that, for example, if the model predicts an article as having a left-leaning media bias, it also provides rationales and justifications for why the article is categorized as leaning left. An example of two versions of explanations is shown in Figure \ref{fig:interface}, illustrating the reasoning behind the GPT classification label of a particular article.

To design the survey experiment, we manually selected 20 recent headline roundups, each containing three news articles, one from the left, center, and right, for a total of 60 recent articles from our dataset. These articles included balanced GPT-predicted bias labels and two types of GPT-generated explanations from previous analyses, which were part of the Study I results. 
These 20 sets of roundups cover a range of topics, including elections, public health, education, and technology, with recent publication dates (from May 2024 to July 2024) to maintain participant interest.

The flow of the human study for each participant is shown in Figure \ref{fig:interface}. To balance cognitive load and maintain data quality, each participant has a maximum of 15 minutes to complete all tasks.
Each participant received one set of headline roundups at random, which included three news articles presented in a random order to minimize selection bias. 
For each article, participants were first asked to read and classify it in one of three categories based on their own judgment without any AI assistance. This assignment aligned with the GPT setup, where we presented the title along with a short snippet and the label definitions. Participants were not informed of the study design and were unaware that each article came from a three-label set. We also asked them to rate their confidence level in their decision on a 7-point scale. Subsequently, participants were instructed to review the same article again, along with AI’s prediction and its Brief or Detailed explanation based on their group conditions. Participants were then asked to reclassify the article as before and rate their confidence level again with the AI’s prediction. They were also asked to describe the factors that influenced their agreement or disagreement with the AI’s prediction as a mandatory open question with a minimum of 50 characters. 
This feedback was used to evaluate their confidence in their decisions, satisfaction with the process, and the reasoning behind their final classifications.

Participants in both conditions received the AI label and explanation together. The Brief condition presented the GPT classification label alongside a concise one-sentence explanation, while the Detailed condition presented its GPT classification label alongside an expanded three-point explanation. Because the Brief and Detailed explanations were generated in separate GPT runs, the label was not held constant across conditions: the same article could receive a different GPT label in the Brief and Detailed conditions (see Figure~\ref{fig:interface}). This occurred for 8 of the 60 articles (13.3\%); the remaining 52 articles received the same label in both conditions. Such differences may result from the longer reasoning required to produce the Detailed explanation, which can lead the model to a different classification. Consequently, our experimental design estimates the effect of different AI label--explanation combinations (Brief vs.\ Detailed), rather than the effect of explanation in isolation from the label. The treatment balance checks (Section~\ref{sec:treatment_balance}) show that the two conditions did not differ significantly in AI label distribution or in alignment with publisher-level reference labels, so they are comparable in aggregate, but differences between conditions should be interpreted as associations with the Detailed AI-information condition rather than as a clean effect of explanation detail. The observed increases in confidence and decision switching could result from (a) the substantive content of the explanations themselves, (b) the increased cognitive processing effort required by lengthier explanations, or (c) the combination of label visibility with extended explanation. While this design reflects how AI systems typically present information to users (label with explanation together), future research with isolated conditions (label alone, explanation alone, or explanation without label) would clarify which components drive the observed effects.
Across the 60 articles, each was evaluated approximately six times (M = 6.2, SD = 2.5) due to randomness. In total, 372 individual task assignments have been completed.

\subsubsection{Validation of Reference Labels for Experimental Stimuli}\label{sec:validation_reference_labels}
As a sanity check on publisher-level labels serving as reference points, we conducted independent annotation with two trained researchers. Both annotators independently labeled each article as Left (1), Center (2), or Right (3) using standardized definitions aligned with political bias classification conventions, with items presented in random order to prevent order bias or influence from each other's judgments. Interrater agreement was strong: the two annotators achieved 91.7\% agreement, with Cohen's $\kappa = 0.87$ and ordinal Krippendorff's $\alpha$ = 0.93, both indicating robust reliability. Accuracy compared to publisher-level reference labels was 90.0\% (n=54) for Annotator 1 (Macro F1 = 0.90) and 81.7\% (n=49) for Annotator 2 (Macro F1 = 0.82).

After independent labeling, the annotators discussed every discrepancy. Because this discussion drew on parallel coverage from the same round-up, the post-discussion labels are not fully independent of the source design, and we do not use them as the primary reference. A key calibration mechanism was comparing parallel versions of the same news event covered by different outlets — seeing how different sources framed identical events helped clarify whether framing differences constituted genuine ideological bias or fell within the bounds of neutral reporting. Both annotators over-assigned the Center category relative to publisher labels, indicating that disagreements were driven primarily by differences in the threshold for assigning ideological bias rather than by disagreement about directional bias. Most disagreements were only one label apart (e.g., Left vs. Center or Center vs. Right); extreme disagreements (Left vs. Right) were rare. Agreement was highest on explicitly ideological articles with strong evaluative language, criticism, or advocacy. The most difficult cases involved subtly framed articles that were largely factual but contained selective emphasis or suggestive wording, exactly the type of borderline case where subjective interpretation varies.

These results confirm that political ideology classification is inherently interpretive, with some articles unavoidably ambiguous as to whether subtle framing constitutes meaningful ideological bias. Our best-case accuracy of 90\% is slightly lower than the 97\% reported in prior work on this dataset ~\cite{baly2020we}, which we attribute to differences in annotation scope (shorter news snippets vs. full articles), our use of a distinct, more recent news collection, and inter-annotator variation in threshold judgments. Nevertheless, this level of agreement supports using publisher-level labels as a simple reference convention for this study. Given the high inter-annotator reliability and overall agreement with publisher labels, we treat publisher-level labels as a reasonably robust reference convention while acknowledging documented disagreement concentrated in a subset of borderline Center-vs-biased classifications. We use ``publisher-level reference labels'' throughout to acknowledge this validation evidence while maintaining appropriate epistemic humility about classification certainty. We retain publisher-level labels, rather than the annotators' consensus labels, as the primary reference for all analyses for three reasons. First, publisher-level labels are external to the research team and were fixed before the human study, whereas consensus labels would introduce the research team's own interpretive thresholds; as noted above, both annotators systematically assigned Center more often than the publisher labels. Second, publisher-level labels are the convention used in prior work on this dataset family \citep{baly2020we}, which keeps our results comparable. Third, the high agreement reported above indicates that substituting annotator labels would change the reference category for only a small subset of borderline items. We therefore interpret all alignment-based results as alignment with this convention, and we regard the borderline Center-versus-biased items as ambiguous rather than as clear cases of alignment or divergence.

\subsubsection{Human Study Evaluation Metrics and Analysis}
At the granular level, we first calculated inter-rater reliability (IRR) using Krippendorff’s alpha \citep{marzi2024k} to measure the level of agreement among participants on the same article and condition, as this metric supports multiple raters and adapts to various data types similar to \citet{zhang2018structured}.
We then inspect various aspects of evaluation at an aggregated level, such as alignment with reference labels, decision changes, time efficiency, confidence shifts, and reliance patterns.
Following established approaches in subjective annotation tasks, we measure participants' alignment with publisher-level reference labels and track how this alignment changes after receiving AI explanations. We explicitly avoid claiming these represent objective ground truth, given the inherent subjectivity of bias classification. We track the number of human decision changes, documenting whether participants shift decisions toward or away from AI predictions, and separately examining cases where AI predictions align with versus diverge from reference labels. We assess changes in participants' confidence levels following GPT explanations, recognizing that confidence changes may occur independently of whether decisions align with publisher-level labels. Time efficiency is measured by recording the time taken to complete the labeling task for each news snippet. This approach allows us to examine not only the direct outcomes of decisions but also the nuanced ways users interact with and rely on AI systems—particularly how explanation-induced influence operates in subjective decision contexts. 

To account for the nested structure of our data, we employed mixed-effects regression modeling. For analysis of decision changes, we estimated a mixed-effects logistic regression predicting whether participants switched their decision, including experimental condition (Brief vs. Detailed explanation) and relevant covariates as fixed effects and a participant-level random intercept to account for repeated observations within participants. For analysis of initial decision patterns, we tested whether baseline participant characteristics (political leaning, education level, news-reading frequency, prior AI familiarity, initial confidence) and article topic predicted participants' initial classifications using an ordinal mixed-effects model with a participant-level random intercept. This hierarchical modeling approach accounts for heterogeneity in participants' baseline responsiveness. Because each article was also evaluated by about six participants, we additionally re-estimated the central models with crossed random intercepts for both participants and articles. We applied the same framework to four participant-level outcomes: decision switching, change in confidence, final agreement with the AI, and movement toward the AI recommendation (among tasks in which the participant initially disagreed with the AI). All models include explanation condition, initial confidence (mean-centered), initial human--AI agreement, and AI--reference-label alignment as fixed effects along with political leaning, regular news reading, and prior AI familiarity. Logistic models were used for binary outcomes and a linear model for confidence change; all were estimated with \texttt{lme4} in R using the BOBYQA optimizer. We report the condition estimates in Table~\ref{tab:mixed_outcomes}, complete coefficients, random-effect variances, and convergence information in Appendix~\ref{app:mixed}, and sensitivity analyses in Section~\ref{sec:Human_Decision_Changes}.

Additionally, we conduct a thematic analysis using the inductive approach proposed by \citet{braun2006using} to evaluate user perspectives on AI predictions and explanations and examine the underlying reasoning influencing user decision-making and confidence.

\begin{table}[!htb]
    \small
    \caption{GPT prediction performance on political ideology and comparison with related studies, evaluated using the weighted average of F1-Score (F1), Accuracy (Acc.), Mean Absolute Error (MAE), and Matthews Correlation Coefficient (MCC). Temperature (T) represents a setting in GPT models that controls the balance between consistency and creativity. The full prompt design can be seen in Table \ref{table:prompt_design}.}
    \label{table:full_news_GPT_performance}
    \centering
    \begin{tabular}{m{2.2cm}|m{4.5cm}ccccc}
        \toprule
        \textbf{Dataset} & \textbf{Model} & \textbf{Prompt} & \textbf{F1} & \textbf{Acc.} & \textbf{MAE} & \textbf{MCC} \\ 
        \midrule
        MediaBias  \citep{swati2023inferential} & IC-BAIT \citep{swati2023inferential} & - & 0.45 & 0.47 & - & - \\
        AIIS  \citep{liu2022politics} & POLITICS \citep{liu2022politics} & - &- & 0.75 & - & -\\
        News-February  \citep{aires2020information} & Poll \citep{aires2020information} & - &0.76 & 0.76 & - & -\\
        \midrule
        \multirow{10}{=}{\citet{baly2020we}} & Fine-tuned BERT \citep{baly2020we} & - & 0.48 & 0.51 & 0.510 & - \\ 
        ~ & Fine-tuned BERT with Twitter Bios \citep{baly2020we} & - & 0.64 & 0.72 & 0.290 & - \\ 
        \cline{2-7}
        ~ & GPT 3.5, T = 0 & Prompt A & 0.38 & 0.40 & 0.662 & 0.127 \\ 
        ~& GPT-4o, T = 0 & Prompt A & 0.59 & 0.62 & 0.420 & 0.466 \\ 
        ~& GPT-4o, T = 0.25 & Prompt A & 0.60 & 0.62 & 0.410 & 0.477 \\ 
        ~& GPT-4o, T = 0.5 & Prompt A & 0.60 & 0.62 & 0.413 & 0.478 \\ 
        ~& GPT-4o, T = 0.25 & Prompt B & 0.55 & 0.57 & 0.470 & 0.386 \\ 
        ~& GPT-4o, T = 0.25 & Prompt C & 0.57 & 0.59 & 0.483 & 0.396 \\ 
        ~& GPT-4o, T = 0.25 & Prompt D & 0.69 & 0.68 & 0.340 & 0.543 \\ 
        ~& GPT-4o, T = 0.5, Self-critique & Prompt E & 0.49 & 0.51 & 0.599 & 0.269 \\ 
        \bottomrule
    \end{tabular}
\end{table}

\section{Results}
\label{Results}
We first report the performance of GPT predictions, then present the MTurk study results, including demographics, quantitative findings, and qualitative insights.

\subsection{Study I - GPT Performance on Historical Dataset}
\label{results_Study I - GPT Prediction Performanc}
To begin with, Study I focused on evaluating the prediction performance of GPT models. We initially experimented with GPT predictions using full news article text to assess the ``out-of-the-box'' potential of LLMs to perform this task compared to state-of-the-art supervised learning methods. We have found that GPT demonstrates comparable or improved performance. Table \ref{table:full_news_GPT_performance} presents our GPT experiments using full news articles and compares them to supervised learning methods. 
Except for the MediaBias dataset, sourced from Media Bias Fact Check\footnote{\url{https://mediabiasfactcheck.com/}}, all other datasets in Table \ref{table:full_news_GPT_performance} were originally obtained from AllSides \footnote{\url{https://www.allsides.com}}. The IC-BAIT \citep{swati2023inferential} framework is based on a pre-trained neural network. The POLITICS \citep{liu2022politics} model relies heavily on a pre-trained masked language model focusing on entities and sentiments. The Poll \citep{aires2020information} model uses a strategy with information theory concepts. For a more direct comparison on the same dataset, the fine-tuned BERT \citep{baly2020we} uses a triplet loss pre-training approach and achieves an F1 score of 0.48 without providing any additional information beyond the news article itself. The same model raises the F1 score to 0.64 after including Twitter bios of media followers, but it still underperforms compared to Prompt D (F1 score of 0.69).
To maintain a fair comparison, the LLM was provided with identical input content, excluding both the news title and media source. GPT was instructed to generate only a classification label, without any accompanying explanation.

\begin{table}[!htb]
    \small
    \caption{GPT performance on news article snippets, evaluated using the weighted average of Precision (P), Recall (R), F1-Score (F1), Accuracy (Acc.), Mean Absolute Error (MAE), and Matthews Correlation Coefficient (MCC), across various input and output configurations.}
    \label{table:short_news_performance}
    \centering
    \begin{tabular}{lll|cccccc}
        \toprule
        \textbf{Model} & \textbf{Content Provided} & \textbf{Response Requirement} & \textbf{P} & \textbf{R} & \textbf{F1} & \textbf{Acc} & \textbf{MAE} & \textbf{MCC} \\ 
        \midrule
        GPT-4o & Article 
        & Label & 0.57 & 0.45 & 0.42 & 0.45 & 0.580 & 0.221 \\ 
        GPT-4o mini & Article & Label & 0.58 & 0.41 & 0.38 & 0.41 & 0.650 & 0.138 \\ 
        GPT-4o & Article & Label + Brief explanation & 0.49 & 0.43 & 0.40 & 0.43 & 0.632 & 0.173 \\ 
        GPT-4o mini & Article & Label + Brief explanation & 0.51 & 0.42 & 0.38 & 0.42 & 0.617 & 0.165 \\ 
        GPT-4o & Article + Headline & Label + Brief explanation & 0.48 & 0.42 & 0.39 & 0.42 & 0.640 & 0.162 \\ 
        GPT-4o & Article & Label + Detailed explanation & 0.46 & 0.43 & 0.41 & 0.41 & 0.670 & 0.163 \\ 
        GPT-4o & Article + Headline & Label + Detailed explanation & 0.45 & 0.42 & 0.39 & 0.42 & 0.661 & 0.139 \\ 
            \bottomrule
    \end{tabular}
\end{table}

Analyzing various settings between GPT models, prompts, and \textit{Temperature}, we found that GPT-4o consistently outperforms GPT-3.5. Shorter prompts generally produced better performance among the five different prompt designs (see Table \ref{table:prompt_design}). We observed that including highly detailed indicators or bias definitions within the prompt led to a reduction in the GPT model’s predictive accuracy. As shown in Table \ref{table:full_news_GPT_performance}, Prompt A, with a shorter set of instructions (Acc. = 62\%), outperforms Prompt C, which has more detailed instructions (Acc. = 59\%). We hypothesize that this may occur due to the model constraining itself within the provided list of indicators. Further experimentation with varying levels of detail in the prompts showed minimal differences in accuracy between prompts containing numerous indicators and those limited to well-known and general indicators. However, a near 10\% improvement in accuracy, from below 60\% to an average of 68\%, was observed when the source names of the articles were included in the prompt (i.e., Prompt D). 
Although Prompt D achieved the highest accuracy during our experiments, we believe providing a bias label for the media source may not be the best approach for these tasks. GPT might focus on keywords in the article rather than fully analyzing the news content. Additionally, we tested whether incorporating additional GPT self-evaluations would improve accuracy by using Prompt E. However, we observed only a minimal change, with accuracy increasing from 49\% to 51\%. 
Finally, we observed that increases in segment did not lead to substantial or consistent improvements. In some cases, higher values produced slightly worse results. Therefore, we decided to keep \textit{Temperature} = 0 to ensure reproducibility in future studies.

\subsection{Study I - GPT Experiments Using Our Recent News Dataset} 
We conducted additional experiments using our collected dataset to prepare for the human study and evaluate GPT-generated explanations. Based on prior results, we primarily selected GPT-4o for testing and also assessed GPT-4o mini, a cost-effective model. We set the temperature (\textit{T}) to 0, as higher values did not yield sustained improvements in accuracy. Keeping \textit{T} at 0 ensures consistency and enhances reproducibility of the generated outputs. We used prompt design version A, which demonstrated the best performance in previous tests. For the Brief explanation request, we appended \textit{``And explain your choice in a short sentence''} to prompt design A. For more Detailed explanations, we included: \textit{``In addition, please provide a clear explanation of why the article was classified in that category in 3 bullet points. Focus on the strongest and most obvious indicators, such as the tone and content of the article. Mentions of political figures or organizations. The typical political bias of the publication. Any notable policy stances or beliefs highlighted in the article.''}

Table \ref{table:short_news_performance} presents a comparison of model performance starting with the same setup as for the previous historical dataset. Overall, the GPT-4o mini model shows lower accuracy compared to the GPT-4o. When only the classification label is requested, GPT-4o achieves the highest F1 score of 0.42. Requesting a detailed explanation slightly reduces the F1 score to 0.41, while a Brief explanation results in a further decrease to 0.40. Additionally, including news headlines as supplementary information slightly improves accuracy for Detailed explanations, increasing from 41\% to 42\%. However, this inclusion slightly worsens the results for Brief explanations, decreasing from 43\% to 42\%. 
Based on previous results, GPT-4o achieves 62\% alignment with reference labels when processing full news articles, declining to 42\% on article snippets. This drop likely reflects reduced contextual information in brief texts, making alignment with publisher conventions more difficult. 
Our primary objective is to investigate how participants respond to GPT explanations and whether they maintain trust even when GPT makes errors. Accordingly, we did not conduct additional tuning for shorter news snippets. We included the news headline in GPT predictions and explanations to provide participants with clearer context and facilitate understanding of the content.

\begin{figure}[tb]  
    \centering
    \includegraphics[width=0.96\textwidth]{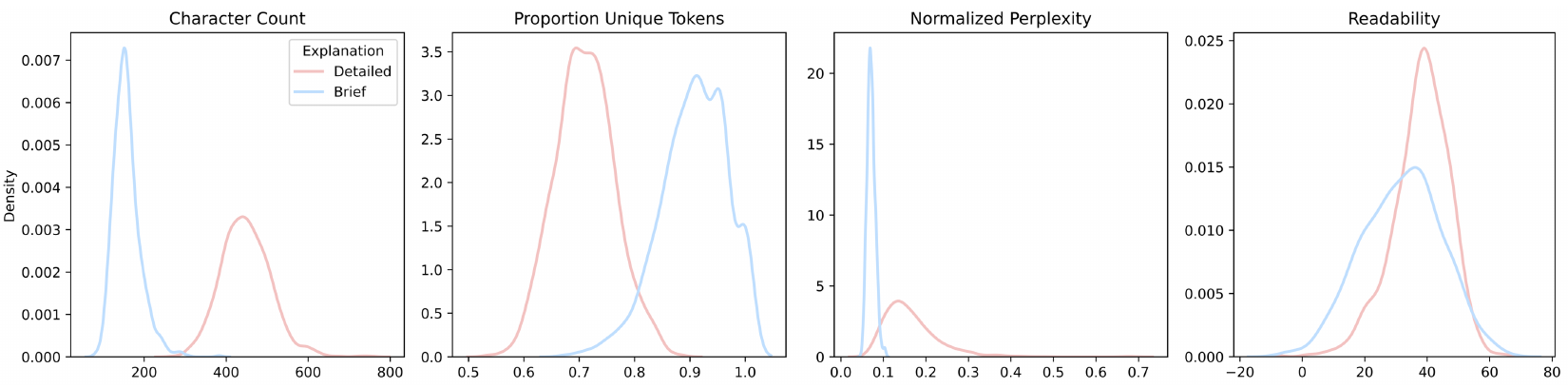}  %
    \caption{Distribution of quality measurements for different types of GPT explanations. The differences in each metric are statistically significant ($p < .001$) using t-tests.}
    \label{fig:text_metrics}
\end{figure}

It is also crucial to evaluate the explanations provided. 
For example, for the same article about Juneteenth, the detailed explanation stated that the article provides “factual information about Juneteenth, its historical significance, and its recognition as a federal holiday,'' includes sources that are ``authoritative and educational,'' and has an ``educational and informative'' tone rather than engaging in ``political commentary or advocacy.'' The Brief explanation similarly emphasized that the article presents factual information about Juneteenth ``without displaying a strong political bias.''
The Brief explanations have an average length of 27 words, while the Detailed explanations are three times longer, with an average of 82 words. Moreover, in Figure \ref{fig:text_metrics}, we compared the character count, proportion of unique tokens, normalized perplexity, and readability using the Flesch reading ease score \citep{flesch1948new}.  Brief explanations show a higher proportion of unique tokens (M = 0.907, SD = 0.060) than Detailed explanations (M = 0.710, SD = 0.053). This suggests that the Brief version is more condensed, which may indicate higher informativeness relative to its length.
The normalized perplexity for Detailed explanations (M = 0.164, SD = 0.065) is higher than for Brief explanations (M = 0.072, SD = 0.009), indicating that Detailed explanations have a more complex structure. Despite this, the readability reveals that Detailed explanations (M = 38.011, SD = 9.119) are easier to read and more comprehensive than Brief explanations (M = 31.912, SD = 12.751). Since Brief explanations consist of only one sentence, we cannot internally evaluate their coherence. For Detailed explanations, the coherence value is 0.90 ± 0.04, indicating that the text is well-organized and the ideas are logically connected.

Additionally, we evaluated the semantic adequacy of LLM-generated explanations using BERTScore~\cite{bert-score}, with the original news articles as the reference. Brief explanations exhibited higher precision (M = 0.860, SD = 0.016) than Detailed explanations (M = 0.842, SD = 0.011), indicating that Brief explanations are more directly aligned with the content of the source articles. In contrast, recall was higher for Detailed explanations (M = 0.830, SD = 0.016) than for Brief explanations (M = 0.821, SD = 0.016), suggesting that Detailed explanations capture a greater proportion of key information from the original news. F1 scores were comparable between conditions, with Brief explanations slightly higher (M = 0.840, SD = 0.014) than Detailed explanations (M = 0.836, SD = 0.014). All differences were statistically significant ($p < .001$), as determined by paired t-tests.
These findings suggest that while Brief explanations offer highly relevant content, likely reflecting summarization or paraphrasing of the source, Detailed explanations convey more comprehensive information, including additional contextual details and the rationale behind the decision generated by the LLM. This pattern highlights a trade-off between conciseness and coverage, emphasizing the need to tailor explanation length to strike a balance between relevance and informativeness in news summarization or decision-support applications.

\subsection{Study II (Crowdsourcing) - Participant Demographics}

We have a fairly balanced distribution in terms of age, political leaning, and gender. Predominantly, the group consists of middle-aged, White, full-time employed individuals with varying political leanings represented across the groups. Full data are shown in Table \ref{tab:dem}.
The majority of participants in both groups expressed disagreement with the use of AI for annotation and labeling prior to this study, with 61.29\% in Group 1 and 56.45\% in Group 2 strongly disagreeing. Some participants in both groups were somewhat disagreeing or were neutral, with a smaller proportion agreeing with AI usage: 14.52\% in Group 1 and 22.58\% in Group 2. The majority of participants in both groups reported being regular news readers, with 87.1\% in Group 1 and 83.87\% in Group 2. A small number in each group indicated they were not regular news readers (12.9\% in Group 1 and 16.13\% in Group 2).

\begin{table}[tb]
\small
\caption{Demographic Information of Participants: 
\textbf{Group 1} consists of participants who were exposed to the Brief version of GPT explanations, while \textbf{Group 2} consists of participants who received the Detailed version of GPT explanations. \textit{Regular News Reader} captures participants' self-identification as regular news consumers rather than an externally validated frequency criterion.}
\label{tab:dem}
\centering
\begin{tabular}{m{4.5cm}ccc}
\toprule
\textbf{Demographic Variable}           & \textbf{Category}                     & \textbf{Group 1}                    & \textbf{Group 2}                     \\  

\midrule
\textbf{Age}                            & Mean ± Std.                           & 47.85 ± 12.42                       & 44.69 ± 10.59                        \\  
\midrule

\multirow{5}{*}{\textbf{Political Leaning}} 
                                        & Democrat                              & 32 (51.61\%)                        & 36 (58.06\%)                         \\ 
                                        & Independent                           & 19 (30.65\%)                        & 11 (17.74\%)                         \\ 
                                        & Republican                            & 8 (12.9\%)                          & 13 (20.97\%)                         \\ 
                                        & Other                                 & 2 (3.23\%)                          & 1 (1.61\%)                           \\ 
                                        & Prefer not to say                     & 1 (1.61\%)                          & 1 (1.61\%)                           \\  
\midrule

\multirow{3}{*}{\textbf{Gender}}        & Female                                & 31 (50\%)                           & 28 (45.16\%)                         \\ 
                                        & Male                                  & 30 (48.39\%)                        & 33 (53.23\%)                         \\ 
                                        & Prefer not to say                     & 1 (1.61\%)                          & 1 (1.61\%)                           \\  
\midrule

\multirow{6}{*}{\textbf{Race}}          & White                                 & 48 (77.42\%)                        & 53 (85.48\%)                         \\ 
                                        & Black or African American             & 7 (11.29\%)                         & 2 (3.23\%)                           \\ 
                                        & Hispanic or Latino                    & 2 (3.23\%)                          & 1 (1.61\%)                           \\ 
                                        & Asian                                 & 2 (3.23\%)                          & 4 (6.45\%)                           \\ 
                                        & Other                                 & 2 (3.23\%)                          & 2 (3.23\%)                           \\ 
                                        & American Indian or Alaska Native       & 1 (1.61\%)                          & 0 (0\%)                              \\  
\midrule
\multirow{5}{*}{\parbox{4.3cm}{\textbf{I have used AI specifically for annotation or labeling tasks before.}}}   
                                        & Strongly disagree                     & 38 (61.29\%)                        & 35 (56.45\%)                         \\ 
                                        & Somewhat disagree                     & 8 (12.9\%)                          & 8 (12.9\%)                           \\ 
                                        & Neither agree nor disagree            & 7 (11.29\%)                         & 5 (8.06\%)                           \\ 
                                        & Somewhat agree                        & 6 (9.68\%)                          & 10 (16.13\%)                         \\ 
                                        & Strongly agree                        & 3 (4.84\%)                          & 4 (6.45\%)                           \\ 
\midrule
\multirow{2}{*}{\textbf{Regular News Reader}} 
                                        & Yes                                   & 54 (87.1\%)                         & 52 (83.87\%)                         \\ 
                                        & No                                    & 8 (12.9\%)                          & 10 (16.13\%)                         \\ 
\bottomrule
\end{tabular}
\end{table}

\subsection{Study II (Crowdsourcing) - Quantitative Results}

In this section, we present key survey findings and examine correlations with demographic factors. We analyze trends in human classifications relative to reference labels and explore how AI explanations affect decision-making and shifts in confidence. Note that the first task refers to classification without AI assistance, and the second task refers to classification with AI explanation.

\subsubsection{Treatment Balance Checks}\label{sec:treatment_balance}

To verify that article and participant characteristics were balanced across explanation conditions, we examined the distribution of key variables. Table~\ref{tab:treatment_balance} presents descriptive statistics for the Brief and Detailed explanation groups. The AI label distribution, AI-reference label alignment, and participants' initial agreement with the AI did not differ significantly between conditions (all $p > .05$), consistent with successful random assignment on these measures. Participants' initial confidence, however, was significantly higher in the Brief condition than in the Detailed condition ($p = .012$), indicating a baseline imbalance on this measure. We note this baseline difference when interpreting subsequent results involving confidence, rather than attributing observed differences in confidence-related outcomes solely to explanation length. Because GPT labels were generated separately for each explanation format and could differ for the same article, the balance in AI label distribution and AI--reference alignment is important: it indicates that the Brief and Detailed conditions were comparable in the labels they presented, even though labels were not fixed at the article level. Article composition was also balanced by design: each participant received one randomly assigned roundup containing one article from each publisher-level category (Left, Center, and Right), so every participant in both conditions evaluated exactly one article per reference category, and both conditions drew from the same 20 roundups.

\begin{table}[!ht]
\small
\caption{Distribution in Brief and Detailed explanation conditions. Differences are expressed as Brief minus Detailed.}
\label{tab:treatment_balance}
\centering
\begin{tabular}{lcccc}
\toprule
\textbf{Variable} & \textbf{Brief} & \textbf{Detailed} & \textbf{Difference} & \textbf{Statistic} \\
\midrule
\multicolumn{5}{l}{\textit{AI Label Distribution (\%)}} \\
\quad Left   & 23.1\% & 29.0\% & $-$5.9 pp & \multirow{3}{*}{$\chi^2$(2) = 2.39, $p$ = .30} \\
\quad Center & 44.6\% & 37.6\% & $+$7.0 pp & \\
\quad Right  & 32.3\% & 33.3\% & $-$1.0 pp & \\
\midrule
AI-Reference Alignment (\%) & 46.8\% & 52.2\% & $-$5.4 pp & $\chi^2$(1) = 0.87, $p$ = .35 \\
Initial Human-AI Agreement (\%) & 63.4\% & 59.1\% & $+$4.3 pp & $\chi^2$(1) = 0.56, $p$ = .46 \\
Final Human-AI Agreement (\%) & 68.3\% & 75.3\% & $-$7.0 pp & $\chi^2$(1) = 1.91, $p$ = .17 \\
\midrule
Initial Confidence, M (SD) & 5.48 (1.26) & 5.11 (1.56) & $+$0.37 & $t$(370) = 2.53, $p$ = .012$^{*}$ \\
Final Confidence, M (SD) & 5.85 (1.23) & 5.60 (1.43) & $+$0.25 & $t$(370) = 1.83, $p$ = .068 \\
\bottomrule
\multicolumn{5}{l}{\footnotesize $^{*}p < .05$}
\end{tabular}
\end{table}

\subsubsection{Time Efficiency}
Upon analyzing the task completion times, we found that the Brief explanation group (Group 1) spent less time on the second task (M = 81.21 seconds, SD = 47.91) compared to the Detailed explanation group (Group 2) (M = 97.07 seconds, SD = 167.51). However, the two-tailed t-test indicates this difference was not statistically significant. This difference is likely due to the Brief explanations (M = 24.77, SD = 4.66) having a lower word count than the longer explanations (M = 53.92, SD = 27.89) in the selected 60 news articles.

\subsubsection{Analysis with the Demographic Factors}
We examined whether demographic factors systematically influenced decision-making patterns, using both bivariate and multivariate approaches. First, we conducted point-biserial correlations to test whether individual differences predicted final classification decisions. Political leaning showed no statistically significant correlation with final classification alignment to reference labels (rpb = -0.09, $p = .081$). Similarly, regular news-reading frequency ($p = .182$) and prior AI experience ($p = .169$) showed no significant associations with final decision patterns.
Overall, descriptive patterns revealed that males tended to favor center-right classifications while females leaned toward center-left classifications. However, this gender difference was not statistically significant ($\chi^2(2) = 1.5, p = 0.472$). Collectively, these demographic analyses indicate that individual characteristics are stable across the sample but do not substantially predict initial classifications or final alignments with reference labels. 

To rigorously test whether demographic characteristics systematically shape initial judgments independent of other factors, we employed an ordinal mixed-effects model with a participant-level random intercept. Consistent with the bivariate correlations, none of the examined characteristics (political leaning, education level, news reading frequency, or prior AI familiarity, etc.) showed statistically significant associations with initial classification decisions (all $p > .05$). This hierarchical analysis confirms that while individual demographic differences exist in our sample, they do not substantially determine how participants initially interpret political news articles on the left-center-right spectrum. Importantly, this finding sets up a critical contrast: while baseline demographic factors fail to predict decision patterns, explanation format substantially influences decision switching and confidence, suggesting that situational factors override individual predispositions in shaping responses to political news classification tasks.

\begin{figure}[!htb]  
    \centering
      \includegraphics[width=0.6\textwidth]{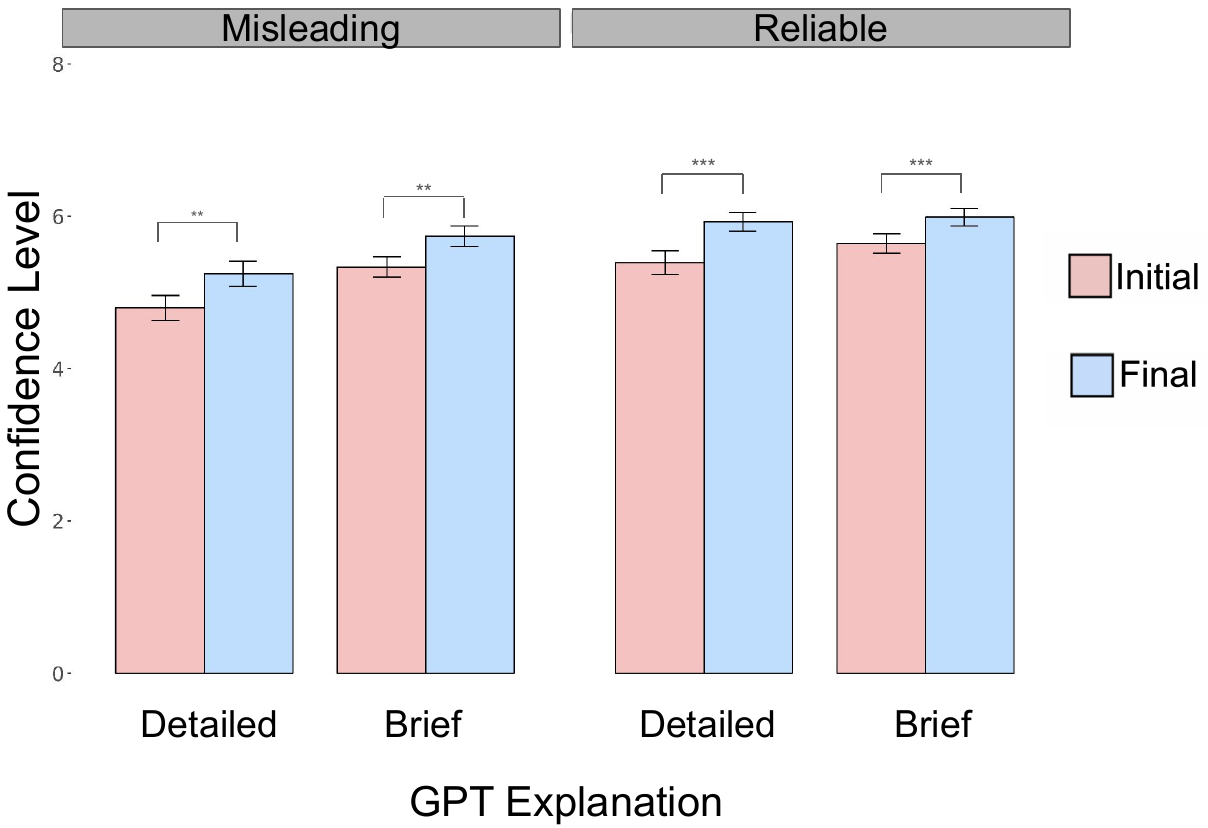}
    \caption{Confidence levels across all conditions for both initial and final decision with Standard Error (SE). `Reliable' corresponds to Aligned and `Misleading' to Divergent. Differences between initial and final in each group are statistically significant, with *** indicating \textit{$p < .001$}, ** indicating \textit{$p < .01$}.}
    \label{fig:Confidence_levels}
\end{figure}

\subsubsection{Confidence Changes}
\begin{figure}[tb]  
    \centering
    \includegraphics[width=0.7\textwidth]{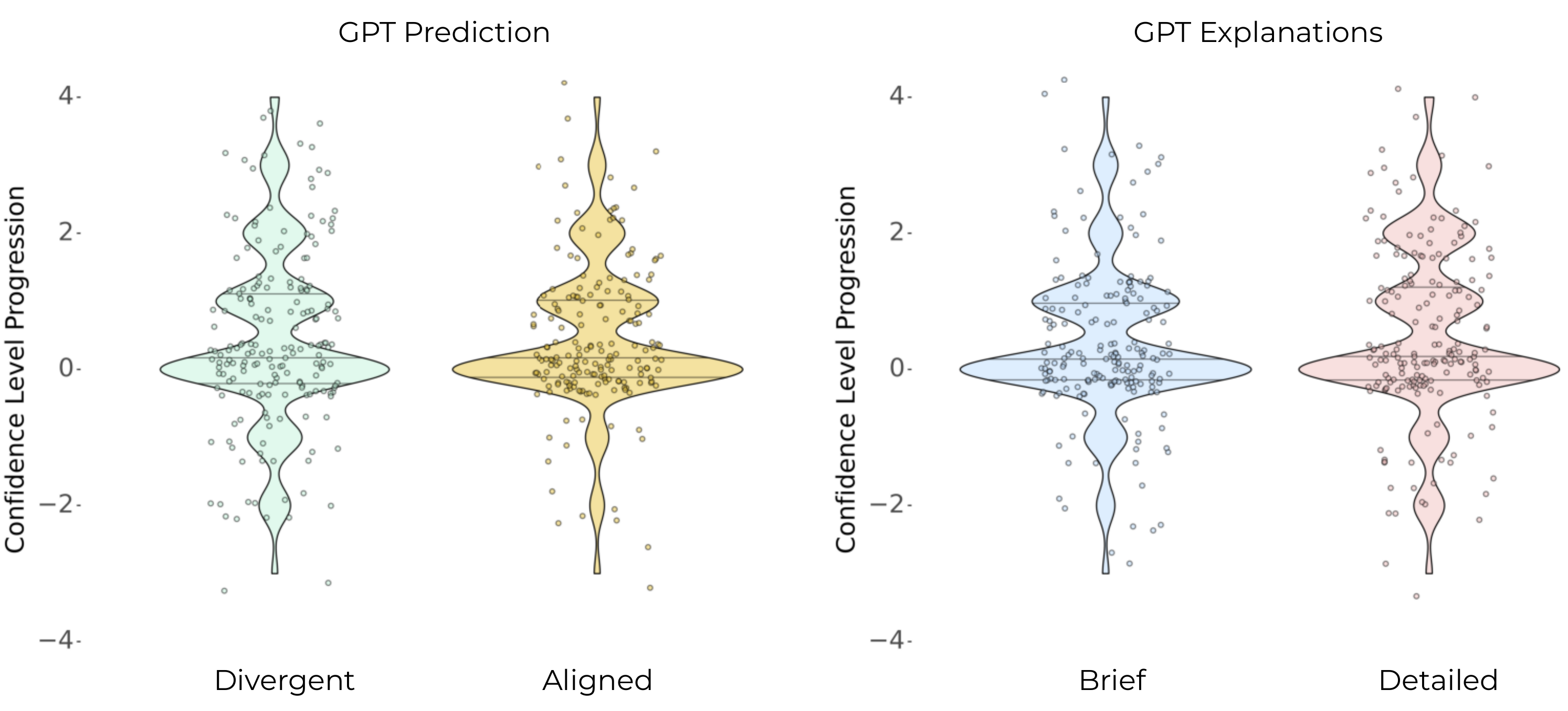}
    \caption{Comparison of Confidence Level Changes for initial and final tasks between different groups.  Confidence levels are measured on a scale of 1 to 7 for both tasks.}
    \label{fig:change_confidence_levels}
\end{figure}

Regarding confidence in labeling decisions, there was an increase in confidence from the initial decision (M = 5.29, SD = 1.44) to the final decision (M = 5.75, SD = 1.34). This difference is statistically significant ($p < .001$) using a paired t-test. 
Figure \ref{fig:Confidence_levels} shows that confidence levels are higher with AI assistance compared to decisions made without it, regardless of whether AI predictions align with or diverge from reference labels or the type of explanation provided. These differences are also statistically significant using paired t-tests. Although initial confidence levels may be low, final confidence levels generally tend to be higher after using AI assistance. Figure \ref{fig:change_confidence_levels} demonstrates that the distributions of changes in confidence levels are similar across different conditions. The Detailed explanation (M = 0.49, SD = 1.27) led to a slightly greater increase in confidence compared to the Brief explanation (M = 0.38, SD = 1.10), though this difference is insignificant using a two-tailed t-test. Additionally, there is no significant difference between the effects of reliable (M = 0.46, SD = 1.10) and misleading (M = 0.45, SD = 1.31) conditions, where AI predictions match or diverge from reference labels. Overall, confidence levels increased across all conditions, indicating that AI assistance generally boosts confidence across conditions.

\subsubsection{Task Performance and Decision Changes}
As shown in Table \ref{table:Performance of decision-makin}, GPT predictions aligned with reference labels in 49\% of the 60 selected articles. When we asked participants to initially classify articles without AI assistance, they aligned with reference labels 47\% of the time. After receiving AI predictions and explanations, the overall alignment remained similar at approximately 50\%, though some participants changed their decisions in response to the explanations. When the AI prediction fully aligned with reference labels, participants' alignment improved from 68\% to 76\%, representing an 8 percentage point increase. When the AI prediction fully diverged from reference labels, participants' alignment dropped from an initial 26\% to 18\%.  The negative MCC value indicates that human performance was worse than random guessing. 
%
%
These results demonstrate the dual nature of AI assistance in this task. When GPT predictions aligned with reference labels and were accompanied by clear reasoning, AI assistance substantially improved participants' alignment with those labels, reducing divergence from those labels by up to 25\%. However, when GPT's predictions diverged, participants often shifted toward the AI's recommendation, even when it diverged from conventional labels. This pattern underscores that participants' responsiveness to AI assistance may be driven substantially by explanation persuasiveness rather than alignment with reference labels.

\begin{table}[!ht]
\small
    \caption{Performance of decision-making with Precision (P), Recall (R), F1-Score (F), Accuracy (Acc.), Mean Absolute Error (MAE), Matthews correlation coefficient (MCC), and inter-rater reliability (IRR), measured by the percentage agreement between human decisions and GPT suggestions. Avg refers to the weighted average. Bold numbers represent the optimal values for their respective evaluations. When using all data points, each class has a support of 124. When considering only decisions where GPT predictions align with publisher-level labels, the support for the left, center, and right classes is 55, 60, and 69, respectively. When focusing only on cases where GPT predictions diverge from reference labels, the support for the left, center, and right classes is 69, 64, and 55, respectively.}
    \label{table:Performance of decision-makin} 
    \centering
    \begin{tabular}{m{5cm}cccccccc}
    \toprule
    \centering \textbf{Performance} & \textbf{Class} & \textbf{P} & \textbf{R} & \textbf{F1} & \textbf{Acc.} & \textbf{MAE} & \textbf{MCC} & \textbf{IRR} \\ 
  \midrule
        \centering \textbf{All Tasks (n = 372)}  & Left & 0.57 & 0.44 & 0.50 & ~ & ~ & ~ \\ 
        \multirow{3}{=}{\centering GPT Decision}  & Center & 0.39 & 0.48 & 0.43 & ~ & ~ & ~ \\ 
        ~ & Right & 0.57 & 0.56 & 0.56 & ~ & ~ & ~ \\ 
        ~ & Avg & 0.51 & 0.49 & 0.50 & 0.49 & 0.589 & 0.244 \\  \cline{2-9} 
        \multirow{4}{=}{\centering Human Initial Decision} & Left & 0.49 & 0.36 & 0.42 & ~ & ~ & ~ \\ 
        ~ & Center & 0.41 & 0.63 & 0.50 & ~ & ~ & ~ \\ 
        ~ & Right & 0.55 & 0.40 & 0.47 & ~ & ~ & ~ \\ 
        ~ & Avg & 0.48 & 0.47 & 0.46 & 0.47 & 0.648 & 0.205 & 61.29\%\\  \cline{2-9} 
        \multirow{4}{=}{\centering Human Final Decision} & Left & 0.51 & 0.35 & 0.41 & ~ & ~ & ~ \\ 
        ~ & Center & 0.41 & 0.63 & 0.50 & ~ & ~ & ~ \\ 
        ~ & Right & 0.54 & 0.42 & 0.47 & ~ & ~ & ~ \\ 
        ~ & Avg & 0.49$\uparrow$ & 0.47 & 0.46 & 0.47 & 0.642$\downarrow$  & 0.205 & 71.77\%$\uparrow$ \\ 
  \midrule
        \textbf{GPT aligns with labels  (n = 184)}  & Left & 0.74 & 0.64 & 0.69 & ~ & ~ & ~ \\ 
        \multirow{3}{=}{\centering Human Initial Decision} & Center & 0.57 & 0.80 & 0.67 & ~ & ~ & ~ \\ 
        ~ & Right & 0.79 & 0.61 & 0.69 & ~ & ~ & ~ \\ 
        ~ & Avg & 0.71 & 0.68 & 0.68 & 0.68 & 0.380 & 0.530 & 67.93\% \\  \cline{2-9} 
        \multirow{4}{=}{\centering Human Final Decision} & Left & 0.85 & 0.73 & 0.78 & ~ & ~ & ~ \\ 
        ~ & Center & 0.65 & 0.87 & 0.74 & ~ & ~ & ~ \\ 
        ~ & Right & 0.84 & 0.70 & 0.76 & ~ & ~ & ~ \\ 
        ~ & Avg & \textbf{0.78}$\uparrow$  & \textbf{0.76}$\uparrow$  & \textbf{0.76}$\uparrow$  & \textbf{0.76}$\uparrow$  & \textbf{0.283}$\downarrow$ & \textbf{0.650}$\downarrow$  & \textbf{76.09\%}$\uparrow$ \\ 
          \midrule
 \textbf{GPT diverges from labels  (n = 188)} & Left & 0.22 & 0.14 & 0.18 & ~ & ~ & ~ \\ 
        \multirow{3}{=}{\centering Human Initial Decision} & Center & 0.29 & 0.47 & 0.36 & ~ & ~ & ~ \\ 
        ~ & Right & 0.21 & 0.15 & 0.17 & ~ & ~ & ~ \\ 
        ~ & Avg & 0.24 & 0.26 & 0.24 & 0.26 & 0.910 & -0.131 & 54.79\% \\  \cline{2-9} 
        \multirow{4}{=}{\centering Human Final Decision} & Left & 0.08 & 0.04 & 0.06 & ~ & ~ & ~ \\ 
        ~ & Center & 0.23 & 0.41 & 0.30 & ~ & ~ & ~ \\ 
        ~ & Right & 0.10 & 0.07 & 0.09 & ~ & ~ & ~ \\ 
        ~ & Avg & 0.14$\downarrow$ & 0.18$\downarrow$ & 0.15$\downarrow$ & 0.18 $\downarrow$& 0.995$\uparrow$ & -0.261$\downarrow$ & 67.55\%$\uparrow$ \\ 
         \bottomrule
    \end{tabular}
\end{table}

\subsection{Study II (Crowdsourcing) - Level of Agreement}
\label{Level of Agreement}
Considering the annotation choices ``Left,'' ``Center,'' and ``Right'' as ordinal values, participants generally exhibited low agreement for initial decisions (M = 0.18, SD = 0.25) across three tasks per news topic. This IRR increased for final decisions (M = 0.28, SD = 0.24), reflecting the influence of GPT in persuading and improving consensus among participants. Some news topics showed strong discrepancies (e.g., the topic ``Hungary Assumes EU Council Presidency, Orbán Visits Kyiv to Push for Ceasefire'' had an IRR of -0.15), while others demonstrated perfect agreement (e.g., ``UN Court Says Israeli West Bank Settlements Violate International Law'' had the highest IRR of 1). In the Brief explanation group, overall IRR improved by 45\%, rising from 0.20 to 0.29. However, the Detailed explanation group showed a greater improvement, with IRR increasing by 60\%, from 0.20 to 0.32. Interestingly, agreement levels in the Brief group remained the same or showed slight improvement. In contrast, the Detailed group had a greater impact, with both sustained positive and negative changes for specific news topics.

In addition, we analyzed the agreement between participants’ choices and GPT suggestions. The match rate between the human initial choice and GPT recommendation serves as the baseline. As shown in Table \ref{table:Performance of decision-makin}, presenting GPT suggestions significantly increases agreement, achieving an average of 71.77\%. When GPT predictions align with reference labels, the agreement rate rises from 67.93\% to a peak of 76.09\%. Notably, even when GPT predictions diverge from reference labels, it still leads to a more substantial 12.76\% increase in agreement. This indicates a strong influence of GPT suggestions on participants' choices; whether such agreement reflects inappropriate reliance cannot be determined without a defensible benchmark.

\begin{table}[tb]
\small
\caption{Human Decision Changes or Remains in Initial and Final Classification of News Tasks Based on GPT Explanation Length and Label Alignment.  ‘Aligned’ and ‘Divergent’  refer to whether the label chosen aligns with or diverges from the publisher-level reference label. For example, presenting Brief AI explanations aligned with the reference label helps people switch from divergent to aligned decisions 4 times. Bold numbers indicate the highest value in each row. The number in the figure represents whether the decision remained unchanged or was altered when presenting GPT predictions divergent from publisher labels (left) and aligned with publisher labels (right). In the bar chart, `Misleading GPT Information' and `Reliable GPT Information' correspond to GPT predictions divergent from and aligned with publisher-level labels, respectively.}

\label{tab:decision_change_table}
    \centering
    \includegraphics[width=0.6\textwidth]{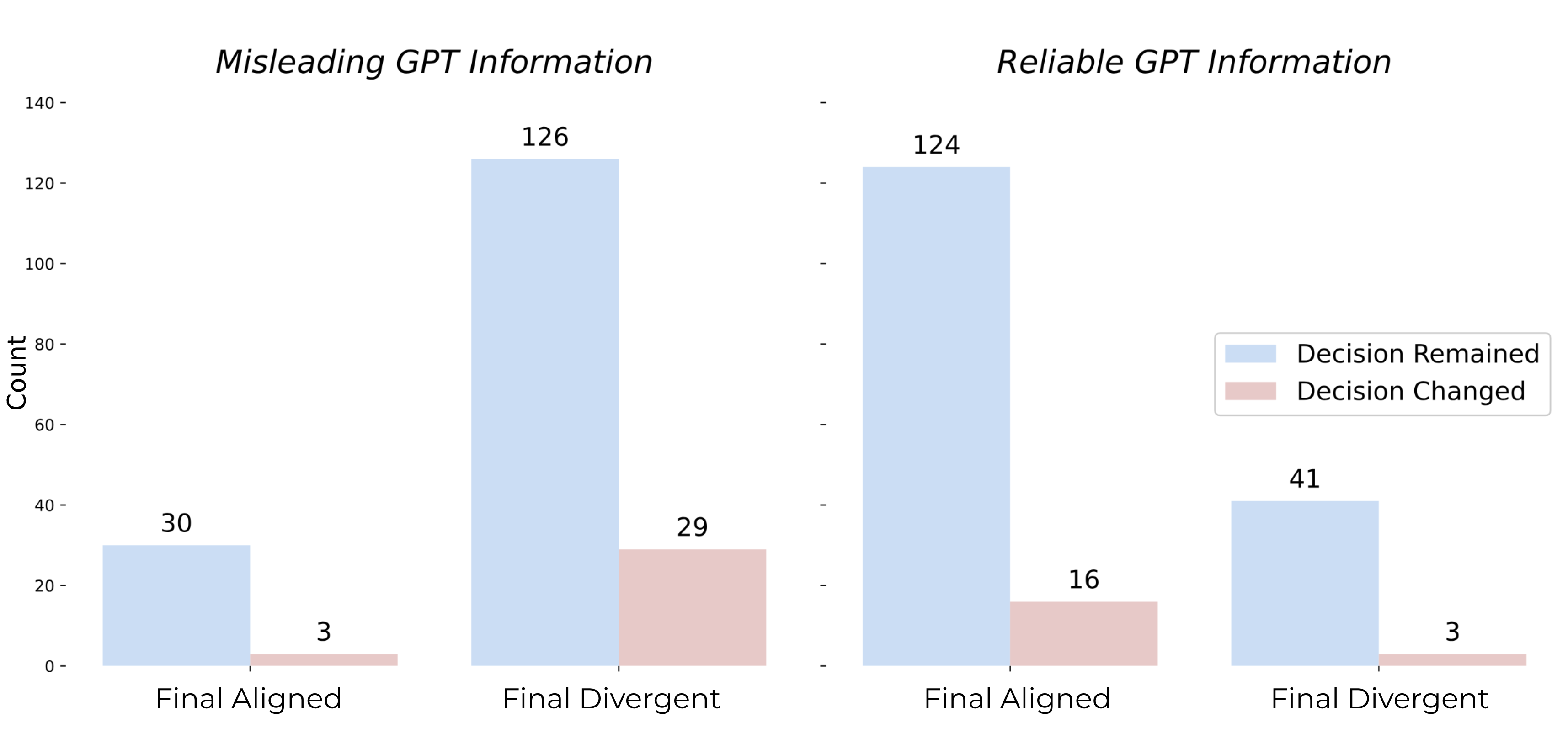}
    \begin{tabular}{m{3cm}m{3.5cm}m{1.5cm}m{1.5cm}m{1.5cm}m{1.5cm}}
     \toprule
         ~ & \textbf{Label Alignment} & \textbf{Aligned \& Brief}  & \textbf{Divergent \& Brief}   & \textbf{Aligned \& Detailed} & \textbf{Divergent \& Detailed} \\ \midrule
        \multirow{3}{=}{\centering \textbf{Decision Changed} \\ n= 51 (14\%)} 
        & Divergent to Aligned & 4 & 1 & \textbf{12} & 2 \\ 
        ~& Aligned to Divergent & 0 & 5 & 1 & \textbf{13} \\ 
        ~& Divergent to Divergent & 1 & 3 & 1 & \textbf{8} \\ 
        \midrule
        \multirow{2}{=}{\centering \textbf{Decision Remained} \\ n = 321 (86\%)}
        & Both Aligned & 56 & 18 & \textbf{68} & 12 \\ 
        ~& Both Divergent & 26 & \textbf{72} & 15 & 54 \\ 
    \bottomrule
    \end{tabular}
\end{table}

\subsection{Study II (Crowdsourcing) - Human Decision Changes}
\label{sec:Human_Decision_Changes}

To investigate the impact of decision changes and explanation-induced influence, we first report aggregate patterns and then present mixed-effects regression results that account for participant-level heterogeneity. Across all conditions, a total of 51 decisions changed out of 372 task assignments, as shown in Table \ref{tab:decision_change_table}. Specifically, 19.89\% (n=37) of decisions changed when Detailed explanations were presented, compared to 7.53\% (n=14) with Brief explanations, suggesting that the Detailed AI label--explanation condition is associated with stronger persuasive effects than the Brief condition.
%
%
When GPT predictions diverged from reference labels, some participants moved away from their initially aligned decisions. This occurred for 13 participants in the Detailed explanation condition and five participants in the Brief explanation condition.
The contingency table shows two main patterns. When GPT predictions aligned with reference labels, participants kept their initially aligned choices 67.39\% of the time (n = 124). They also switched to an aligned final decision 8.70\% of the time (n = 16).
When GPT predictions diverged from reference labels, participants still kept their original aligned labels in 67.02\% of decisions (n = 126). However, 15.43\% of decisions (n = 29) shifted toward the GPT prediction. Another 5.85\% of cases (n = 11) shifted from one non-aligned position to another. This suggests that GPT explanations can influence user decisions even when they do not directly move users from an aligned answer to the GPT-predicted label.

We tested these effects using mixed-effects logistic regression, and the treatment effect remained statistically significant after adjustment for participant-level heterogeneity. Participants in the Detailed-explanation condition showed significantly higher odds of switching their decision compared to those in the Brief-explanation condition (b = 1.134, 95\% CI [0.334, 1.934], OR = 3.11, p < 0.01). This indicates that the Detailed condition is associated with three times higher odds of decision switching relative to the Brief condition, noting that the two conditions differ in both explanation length and, for some articles, the GPT label shown. To account for article-level dependence, we re-estimated this model with crossed random intercepts for participants and articles, including initial confidence, initial human--AI agreement, and AI--reference-label alignment as covariates (Table~\ref{tab:mixed_outcomes}). The article-level random-intercept variance was estimated at zero, and adding it did not improve model fit ($\chi^2(1) = 0.00$, $p > .99$). The Detailed condition remained associated with higher odds of switching (OR = 3.23, 95\% CI [1.47, 7.11], $p = .004$). The same framework showed that the Detailed condition was associated with higher odds of final agreement with the AI after adjusting for initial agreement (OR = 3.29, 95\% CI [1.30, 8.36], $p = .012$), even though the unadjusted difference in final agreement was not significant (Table~\ref{tab:treatment_balance}). Among tasks in which participants initially disagreed with the AI, the Detailed condition was associated with higher odds of moving to the AI's label (32 of 76 vs.\ 10 of 68; OR = 4.37, 95\% CI [1.62, 11.79], $p = .004$). Change in confidence did not differ between conditions ($b = -0.02$, 95\% CI [$-0.25$, 0.20], $p = .84$). Because only 51 decisions changed, we conducted several sensitivity analyses of the switching result. The condition effect remained significant when excluding the eight articles whose GPT label differed between conditions (OR = 2.94, 95\% CI [1.27, 6.79], $p = .012$), in a model with condition as the only fixed effect (OR = 3.05, 95\% CI [1.59, 5.86], $p = .001$), and with article fixed effects instead of an article random intercept (OR = 7.07, 95\% CI [2.10, 23.84], $p = .002$). In a leave-one-article-out analysis, the condition coefficient ranged from 0.96 to 1.43 and remained significant in all 60 runs (all $p \leq .016$). A Fisher's exact test on the raw switching counts (37 of 186 vs.\ 14 of 186) gave a similar odds ratio (OR = 3.05, $p < .001$). Given the small number of switches (51 total), the confidence intervals are wide, and these effect estimates should be interpreted as preliminary findings that warrant replication in larger samples. The substantial effect of explanation format on decision switching is consistent with the interpretation that explanation-induced influence contributes to decision changes in subjective annotation tasks.

\begin{table}[!htb]
\small
\caption{Mixed-effects estimates of the Detailed (vs.\ Brief) AI label--explanation condition for four participant-level outcomes, with a participant-level random intercept only and with crossed participant and article random intercepts. All models adjust for initial confidence, initial human--AI agreement (except the movement model, where it is constant), AI--reference-label alignment, political leaning, regular news reading, and prior AI familiarity. Estimates are odds ratios for binary outcomes and unstandardized coefficients ($b$) for confidence change, with Wald 95\% confidence intervals. $\sigma^2_{P}$ and $\sigma^2_{A}$ denote participant and article random-intercept variances. Complete coefficients are reported in Appendix~\ref{app:mixed}.}
\label{tab:mixed_outcomes}
\centering
\begin{tabular}{llcccc}
\toprule
\textbf{Outcome} & \textbf{Random effects} & \textbf{Estimate [95\% CI]} & \textbf{$p$} & \textbf{$\sigma^2_{P}$} & \textbf{$\sigma^2_{A}$} \\
\midrule
Decision switching ($n = 372$) & Participant & OR 3.23 [1.39, 7.49] & .006 & 0.141 & -- \\
 & Participant + Article & OR 3.23 [1.47, 7.11] & .004 & 0.141 & 0.000 \\
Final agreement with AI ($n = 372$) & Participant & OR 3.29 [1.24, 8.74] & .017 & 0.668 & -- \\
 & Participant + Article & OR 3.29 [1.30, 8.36] & .012 & 0.668 & 0.000 \\
Movement toward AI ($n = 144$) & Participant & OR 4.37 [1.46, 13.09] & .009 & 0.627 & -- \\
 & Participant + Article & OR 4.37 [1.62, 11.79] & .004 & 0.627 & 0.000 \\
Confidence change ($n = 372$) & Participant & $b = -0.02$ [$-0.25$, 0.20] & .835 & 0.192 & -- \\
 & Participant + Article & $b = -0.02$ [$-0.25$, 0.20] & .835 & 0.192 & 0.000 \\
\bottomrule
\end{tabular}
\end{table}

%


\begin{table}[tb]
    \small
    \caption{Codebook used for thematic analysis to examine perceptions of AI decisions. The last column reports the number of instances labeled with each code. }
    \label{tab: codes}
    \centering
    \begin{tabular}{m{1.5cm}m{1.5cm}m{10.2cm}m{0.5cm}}
        \toprule
        \textbf{Themes} & \textbf{Codes} & \textbf{Definition} & \textbf{\#} \\ 

        \midrule
        
        \multirow{4}{=}{Perceptions or Reactions to AI Decisions} & Agree with AI Decision & Instance where participants agree with the AI’s decision regarding the news article. They perceive the AI’s judgment as accurate. \textit{\textbf{e.g.,} ``The AI position follows what I feel and made sense to me.''} & 131 \\ 
        \cline{2-4}
        ~ & Disagree with AI Decision & Instance where participants disagree with the AI’s decision regarding the news article. They perceive the AI’s judgment as inaccurate. \textit{\textbf{e.g.,} ``I understand the AI’s reasoning, but I disagree with the conclusion...'' } & 32 \\ 
        \cline{2-4}
        ~ & Neutral or Mixed Reactions & Instance where participants express mixed feelings or do not strongly agree or disagree with the AI’s decision. Their response indicates uncertainty or a balanced perspective. \textit{\textbf{e.g.,} ``I can somewhat see why the AI has reasoned the way it did, but I don’t know if I agree based only on the piece of the article that way provided.''} & 23 \\ 
        \cline{2-4}
        ~ & No Reaction / Not Applicable & Instance where participants provide no feedback about AI reasoning, or their response is not relevant to the AI’s decision. \textit{\textbf{e.g.,} ``I think the article actually stuck with the facts instead of making excuses.''} & 186 \\ 
        \midrule
        \multirow{6}{=}{Reasoning} & Expressing Similar Reasoning with AI & Instance where participants express similar reasoning aligned with AI.  \textit{\textbf{e.g.,} ``The points that it’s making about the contrast is exactly how I assessed it.''} & 119 \\ 
        \cline{2-4}
        ~ & Balanced Views & Instance where participants discuss balanced views from either AI or itself, such as pros and cons of the news articles. \textit{\textbf{e.g.,} ``I agree I didn’t see anything that makes me believe the article was left or right.'' } & 31 \\ 
        \cline{2-4}
        ~ & Evidence of Article Bias & When participants notice and point out that the article shows favoritism or prejudice toward a certain viewpoint or group. \textit{\textbf{e.g.,} ``Feeling bad about Biden’s performance is definitely left-leaning....Also calling the Supreme Court tainted is definitely left-leaning.''} & 81 \\ 
        \cline{2-4}
        ~ & Factual Reporting / Accuracy & Instance where participants acknowledge that the article is purely factual without any bias. \textit{\textbf{e.g.,} ``The wording ``horrors of slavery'' made me first think it was biased toward the left, but after rereading it, I think it is mostly objective.''} & 32 \\ 
        \cline{2-4}
        ~ & Difficulty in Assessing Bias & Instance where participants express uncertainty or concerns in assessing the bias in the article.  \textit{\textbf{e.g.,} ``I was on the fence initially about whether it was slanted enough to consider it biased...''} & 13 \\ 
        \cline{2-4}
        ~ & Clarifying Complex News & Instances where participants note the positive impact of AI in helping them better understand the news.  \textit{\textbf{e.g.,} ``Now that I see its perspective, the reasoning makes sense.'' } & 13 \\ 
        \bottomrule
    \end{tabular}
\end{table}

\subsection{Study II (Crowdsourcing) - Qualitative Results}
\label{Qualitative}

We analyzed 372 open-ended responses to \textit{``Why do you agree or disagree with AI?''} After quality checks (Section \ref{section:quality}), two researchers reviewed a sample to develop a preliminary codebook (Table \ref{tab: codes}). The final codebook includes AI reaction categories and subcategories for participants’ reasoning. Coding disagreements were resolved with input from two additional researchers.

\subsubsection{Agreement with AI Decision}
Our analysis indicates a strong alignment (131 instances, 35\%) between participants' decisions and the AI predictions.
For those agreeing with the AI left-leaning predictions, the predominant reasons include perceived critical tones towards conservative figures and positive portrayals of leftist policies. Specifically, 45 instances indicated that the AI’s predictions matched their views. This alignment was due to several factors: the articles were critical of right-wing figures or policies, had a negative tone towards conservatives, or positively portrayed leftist policies. Phrases like \textit{``radical left''} were frequently mentioned in participant responses. One shared \textit{``The AI and I agree that the tone is heavily positive for a leftist policy.''}
In agreement with the center explanations, 44 instances where participants agreed highlighted the articles' factual and neutral presentation. They noted the absence of overt bias and felt that the articles presented information without showing a clear political slant, with one participant saying, \textit{``I agree with the AI because I did not find that this article had any political bias. There was no mention of politicians or politics.'' } 
For right-leaning explanations, 42 instances where participants aligned with the AI reasonings. A critical tone towards leftist figures and patriotic themes are seen as indicators of right-leaning bias. Participants in this category often referenced language or themes that aligned with conservative viewpoints and criticized leftist policies or figures.
Overall, our results reveal that participants’ reactions for agreeing with the AI’s predictions are mostly influenced by their interpretations of the articles’ tones and content in relation to political biases.

\subsubsection{Disagreement with AI Decision}
\label{disagreement}
In analyzing participants' responses to the AI predictions for news articles, participants disagreed with the AI in only 32 (8\%) instances.
For articles categorized as left-leaning, participants frequently disagreed with the AI conclusions. Several instances (n=13) raised concerns about the AI inability to detect subtle nuances or right-leaning bias that they perceived in the articles. For example, 
one participant highlighted the AI’s failure to pick up on deeper nuances, saying that \textit{``the central allegation was basically whitewashed/obfuscated as it focused too much on the clickbait-ish headline."} A few cases (n=3) also believed the AI predictions reflected a slight Republican bias, which they felt unfairly portrayed Democrats in a negative light. Additionally, one instance critiqued the AI’s assessment of tone.
For center-leaning articles, eight instances expressed similar dissatisfaction with the AI’s predictions. They argued that the AI failed to account for nuanced reporting, which they felt was neutral rather than biased. One example pointed out, \textit{``I disagree with AI. Although it does speak about Biden's policy that aligns with the left's values, it does so in a manner of just reporting factual news.''}
Several instances (around seven) emphasized that the AI was overly focused on factual reporting and policy rather than on tone or the implications of how the news was presented. One individual mentioned, \textit{``AI doesn't understand nuance. It is focusing on the policy being discussed, not how it is being discussed.''}
In general, participants across categories raised concerns about the AI's ability to detect subtle biases and nuances. Many felt the AI’s interpretation missed key contextual details, misrepresented factual reporting as biased, or failed to assess the tone accurately. 

\subsubsection{Neutral or Mixed Reactions}
\label{metural_reactions} 
Twenty-three instances (6\%) expressed neutral or mixed reactions to the AI news bias classification. These respondents often felt uncertain about the political leanings of the articles or acknowledged both sides without fully committing to one perspective. This ambivalence suggests a nuanced perception of the AI analysis, where participants neither strongly agreed nor disagreed with the predictions.
Eight instances highlighted the difficulty of determining bias from limited excerpts, with one participant stating, \textit{``I think it is difficult to tell with much confidence without reading more of the article.''} These statements indicate that the provided excerpts were insufficient for a clear judgment, reinforcing a neutral stance.
Another eight participants acknowledged the AI reasoning but remained skeptical or uncertain. One noted, \textit{``The AI is partially right, but also partially wrong because the person talked about Juneteenth losing flavor due to multiculturalism,''} showing partial agreement without full conviction.
Seven responses highlighted the balanced nature of the articles or AI conclusions. For example, one participant said, \textit{``There’s no lean in the article, whether it be left or right, only facts are listed,''} suggesting the content was largely objective.
Overall, these mixed responses reflect careful consideration of AI predictions, with participants recognizing both strengths and limitations. Many saw merit in the AI analysis but refrained from definitive judgments due to article balance or lack of conclusive evidence.

\subsubsection{No Reaction}
In the remaining 186 instances, participants typically did not react strongly or provide perspective regarding the AI’s decisions and explanations. Many participants viewed the content as neutral because it reported factual information, such as court decisions or market trends.
Even when articles have critical remarks about political figures, participants felt the overall presentation remained factual. This is evident from comments like, \textit{``The article is just talking about what happened outside of the courtroom and quoting the person they spoke to...''}, Furthermore, some participants appreciated the inclusion of multiple perspectives, considering it to contribute to a balanced portrayal, as reflected in statements like, \textit{``Seems like a similarly balanced article showing alternative and reasonable reasons for Biden's actions.''} Overall, the combination of factual reporting, balanced language, and inclusion of multiple perspectives led to attitudes of indifference toward AI decisions.

\subsubsection{In-Depth Reasoning Analysis}
\label{indept_analysis}
While the previous subsections aggregated response patterns (35\% agreement, 8\% disagreement, 6\% neutral), this analysis examines the specific reasoning mechanisms underlying each reaction type. Understanding not just \textit{whether} but \textit{how and why} participants agreed, disagreed, or remained uncertain provides insight into the psychological processes through which AI explanations influence subjective judgment. By identifying the thematic patterns in participant reasoning—whether they reference tone, factuality, or interpretive gaps—we can clarify which aspects of explanations drive the observed confidence and agreement effects, directly addressing RQ3 and complementing the quantitative findings on explanation-induced influence for RQ2. This fine-grained understanding informs our design recommendations by showing which explanation features support genuine interpretive reasoning versus which might amplify inappropriate anchoring.

Here, we examine reasons for agreement, disagreement, or neutrality across 372 articles level, focusing on sub-codes and context for deeper insight.

\textbf{Expressing Similar Reasoning with AI.}
Participants found the AI rationale similar to their own in 119 (41\%) cases. In 52 instances, participants aligned with the AI analysis of the tone and bias, stating, \textit{``The AI position follows what I feel and made sense to me.''} In 30 cases, they saw the AI’s evaluation as a fair and factual representation, as one participant commented, \textit{``The article is just describing a holiday.''} Additionally, 22 instances involved recognizing the AI’s agreement with the categorization of political leanings. Lastly, 15 participants had high confidence in the AI’s judgments, reflecting, \textit{``I would mostly agree with the AI predictions for similar reasons.''} Overall, participants often felt that the AI rationale mirrored their own perspectives.

\textbf{Balanced Views.}
31(11\%) instances mentioned balanced views as their reason for aligning with the AI’s evaluation. They described the articles as neutral, factual, or presenting both sides equally, often commenting that the articles did not show a clear bias towards any side. Example include,
\textit{``the language used is very neutral.''} Participants felt that the AI rationale matched their own perception of the articles being balanced and unbiased.

\textbf{Evidence of Article Bias.}
81 (28\%) instances confidently identified article bias and provided reasoning for their judgments. Participants identified that articles demonstrating a left-leaning bias often included criticism of Trump or support for Biden, used language favoring liberal policies, and highlighted progressive issues. Conversely, right-leaning articles typically critiqued Biden or supported conservative viewpoints, used language favoring traditional values, and defended conservative positions while challenging liberal stances.

\textbf{Factual Reporting/ Accuracy.}
Participants generally agreed in 32 cases (11\%) that articles focusing on factual data were perceived as unbiased, often supporting the AI’s predictions about these articles presenting data or historical events without political bias. As one participant stated: \textit{``I am confident the AI predictions are accurate because the article reports real market trends based on facts without showing political bias.''} Reactions to the AI’s bias predictions were more mixed. Most agreed with the AI on center articles, but opinions varied on those labeled as right- or left-leaning. For example: \textit{``I disagree with the AI. Although the article discusses Biden’s policy, it just reports factual news without political bias.''} Thus, interpretations of articles perceived as neutral or fact-based can differ based on individual perspectives, with a focus on the tone rather than the content itself.

\textbf{Difficulty in Assessing Bias.}
In 13 (4\%) instances, participants discussed challenges in identifying bias. For left-leaning articles, some struggled with determining bias, questioning if tone alone was a reliable measure. Participants' perspectives on bias classification varied significantly. For instance, some noted differences in how articles were framed, while others emphasized the difficulty of determining whether tone alone constitutes reliable evidence of bias. Responses on center bias were mixed; some agreed with the AI but were uncertain about its reasoning. Opinions on right-leaning articles also varied. These comments highlight the difficulties in assessing bias consistently and the inherent subjectivity of political ideology classification.

\textbf{Clarifying Complex News.}
At first, the influence of personal biases and limited information on participants' judgments was explicitly seen in 13 (4\%) instances. Participants initially struggled with their own biases or a lack of understanding, which affected their assessments of news articles. However, after further reflection and review of the AI predictions, many participants found that they agreed with the AI assessments. For example, one participant said, \textit{``I feel the AI can take an unbiased view of the article which I feel I am not able to do because of my feelings about Trump.''} This feedback suggests that GPT explanations may help clarify complex news articles that participants found challenging and reduce their biases.



\section{Discussion}
\label{Discussion}
Here, we address our research questions, implications of our findings, and ethical considerations, and outline limitations with future work. 

\subsection{Evaluating LLMs and Human Performance in Media Bias Classification} 
To address \textbf{RQ1}, Study I demonstrates that using GPT-4 to identify biased news is both feasible and effective. Its performance is comparable to leading supervised learning models. Specifically, using full-text news articles, GPT-4 achieves up to 68\% accuracy with an F1 score of 0.69, while a supervised, fine-tuned BERT model reaches 72\% accuracy and an F1 score of 0.64 when additional information is included. Given BERT’s limitations in generalizability and explainability, these results highlight GPT-4’s potential for classifying news bias and supporting efforts to identify media bias more broadly.
We collected more recent news articles as a new dataset to optimize participant engagement in Study II. While the full article links and metadata were recorded, we formatted snippets for this part of the experiment. This abbreviated content was used as microtasks for each participant to maximize task completion on the MTurk platform. In addition, choosing shorter articles over long-form was driven by considerations of cognitive load, engagement, and the ability to assess bias accurately without unnecessary distractions. Our analysis shows that participants could make informed judgments based on the content provided, with only a few mentioning difficulty forming a strong opinion due to limited context. This suggests that the short articles generally allowed participants to form a judgment, although access to full articles might change participants' judgments and confidence (see Section~\ref{sec:limitations}).
GPT achieves 42\% accuracy on article snippets from our recent dataset. This moderate performance level reflects the inherent difficulty of subjective political classification and provides a realistic context for examining how users respond to AI explanations when predictions sometimes diverge from conventional labels.

To answer \textbf{RQ2}, we found that AI assistance significantly increases human self-reported confidence levels in their labeling decisions, regardless of whether AI predictions align with reference labels or the type of explanation provided, as indicated in Figures \ref{fig:Confidence_levels} and \ref{fig:change_confidence_levels}. While Detailed explanations are more likely to influence decision-making, with nearly 1 in 5 decisions being swayed, this may raise concerns that participants could develop over-reliance, similar to the findings in \citet{bansal2021does}. 
As detailed in Section \ref{sec:Human_Decision_Changes}, the effect persisted after accounting for individual differences in baseline characteristics and for article-level variation. This finding underscores a fundamental challenge in designing AI systems for subjective decision-making contexts: explanation quality must be carefully balanced against potential risks of anchoring and over-reliance, which our design can suggest but not directly measure.

Although AI support may be beneficial, human input remains essential. Our Study I showed that providing GPT with full news articles achieves high performance, with the potential for further improvement as LLMs advance. Human involvement may still be necessary even with high-performing AI systems, particularly in high-stakes decision contexts. For instance, in medical health, prior work has mentioned potential drawbacks of over-relying on LLMs, such as the degradation of clinical skills and the shift from human-generated to AI-generated training data \citep{choudhury2024large}. 
This highlights the importance of maintaining and strengthening clinical expertise rather than gradually delegating judgment authority to AI systems. Similarly, while AI could assist with understanding complex policy issues, humans remain essential for making value-based decisions that affect public welfare.

\subsection{Explanation Induced Influence and User Interpretations}
To answer \textbf{RQ3}, our qualitative findings in section \ref{Qualitative} reveal that user perceptions of AI assistants are influenced by the clarity and alignment with their views. Transparent AI assessments often reinforced users' judgments and helped users reconsider their prejudgments, especially when articles appeared neutral. The effectiveness of AI depends on the quality of the explanations and the complexity of the content. Although political inclinations had little impact on decisions, our analysis found that AI explanations with explicit bias evidence in the article were more persuasive and had a stronger influence on shaping user decisions.


Although one study \citep{wang2021explanations} suggests that the effectiveness of AI explanations varies with expertise, we found no strong correlation between expertise (e.g., being regular news readers) and participants' final alignment with reference labels. Notably, in contrast to prior work \citep{wang2023watch} which found reduced trust from limited-knowledge groups, the explanations in our study increased confidence levels across all participant groups uniformly. This finding is consistent with the dual-effect mechanism where explanations boost confidence independent of alignment with reference labels.
This pattern requires careful interpretation. Participants aligned with AI explanations across conditions, and their confidence increased even when AI predictions diverged from reference labels. This suggests that explanations can act as persuasive mechanisms rather than merely supporting balanced judgment. Although AI may prompt reflection in complex political contexts, explanations that persuade without external validation may create risks of unwarranted confidence and misalignment. These findings highlight the need for transparent design that helps users calibrate confidence and maintain reflective judgment rather than defer to AI authority  (see the specific mechanisms in Section~\ref{sec:design_recommendations}).

Furthermore, our GPT explanation was neither designed nor intended to persuade users to align with the AI's annotation. Unlike studies \citep{dragoni2020explainable, singh2024measuring, karinshak2023working}, which focus on designing AI to deliberately influence user behavior or actions, our goal was not to guide users toward a particular conclusion. However, explanation presented during the experiments may inherently carry some persuasive elements or language.  
In our study, 35\% of participants’ comments expressed alignment with the AI’s reasoning, whereas only 8\% expressed strong disagreement. As discussed in Section \ref{Level of Agreement}, GPT’s influence on users’ decisions extends beyond whether its recommendations align with reference labels. These findings suggest that users may align with GPT-generated decisions not only because they perceive them as correct, but also because the explanations themselves shape how users interpret and evaluate the task.


\subsection{Design Recommendations}\label{sec:design_recommendations}
While our empirical study focused on political bias classification in news, the dual-effect findings generalize to any subjective classification domain where multiple valid interpretations exist. Examples include content moderation decisions, medical diagnosis support, legal document classification, and product or complaint categorization. A core tension exists: explanations improve interpretability while potentially increasing inappropriate reliance. This tension appears fundamental to how humans interact with AI reasoning, regardless of the domain. Therefore, we present design principles applicable across these subjective tasks, though implementation details will vary by context and application.

Based on our findings on the dual effect of AI explanations, where explanations provide interpretive support while potentially increasing unwarranted confidence and reliance, we propose design principles for AI systems in subjective domains. Rather than focusing only on explanation quality, designers should recognize and actively manage the confidence and anchoring effects that even well-designed explanations can produce. We organize these recommendations by intervention level, clarifying which require system-level design changes, which can be achieved through user interaction design or prompt engineering, and which depend on task-specific implementation choices.

\subsubsection{System-Level Design Recommendations}
Here, we describe system-level design recommendations.

\textit{Nuance and Tone Recognition.} A common concern among participants was the AI's inability to recognize subtleties in articles, leading to conclusions about bias labels that diverged from publisher labels (Section \ref{indept_analysis}). Developers should implement semantic and sentiment analysis components to improve detection of tone, emotional language, and framing cues. Specific implementation: train or integrate classifiers that detect hedging language (``potentially,'' ``suggests''), emotional indicators (``incredibly,'' ``outrageously''), and contrast language against domain-specific neutral baselines. This improves consistency with reference standards and reduces confident misclassification.

\textit{Contextual Retrieval and Historical Context.}  Rather than generic recommendations, systems should automatically surface: (1) three to five historically similar articles classified differently, showing how identical events received different framings by different sources; (2) fact-check annotations for claims mentioned in the article; and (3) publication guidelines or editorial standards for the source publication, if available. This allows users to calibrate against multiple reference points rather than the AI's single judgment. A lighter-weight version can be achieved through prompting (e.g., asking the AI to describe relevant historical context or related coverage), but automatically retrieving differently classified coverage and fact-check annotations requires retrieval infrastructure at the system level.

\textit{Confidence Score Integration.} AI systems should compute and display a confidence or uncertainty score (e.g., 0.45-1.0) alongside the primary classification. Scores should reflect the model's statistical confidence in the label, not just its output probability. When confidence is below 0.65, the system should flag the prediction as borderline and explicitly recommend human review or multiple interpretations to reduce anchoring on a weakly supported prediction.

\subsubsection{User-Interaction and Prompt-Based Recommendations}
Here, we present user-interaction and prompt-based recommendations.

\textit{Transparent and Justifiable Explanations.} Several participants found the AI's reasoning unclear (Sections \ref{disagreement} and \ref{metural_reactions}). Rather than redesigning the model, practitioners can improve explanations through prompt engineering: (1) instruct the AI to map each reasoning point directly to quoted text or specific text sections (e.g., ``[QUOTE: 'inflammatory language here'] indicates Left-leaning framing''); (2) request the AI to provide counterfactual examples (``If this phrase were changed to [X], the classification would be Center''); (3) ask the AI to acknowledge competing interpretations (``This could also be read as Center because...''); (4) implement a layered presentation where users first see a one-sentence summary, then can expand to three-point summary, then to full reasoning with text excerpts.

\textit{Balanced Perspective Presentation.} Without retraining the model, practitioners can prompt the AI to provide multiple perspectives simultaneously. For example: ``Explain why this article might be classified as Left. Explain why it might be classified as Center. Explain why it might be classified as Right. Which evidence is strongest?'' This mitigates one-sided explanations and helps users see where interpretative uncertainty lies. Alternatively, run the same article through multiple prompt framings and surface the diversity of possible classifications.

\subsubsection{Task-Design Level Recommendations}
Here, we present task-design level recommendations.

\textit{Calibration and Confidence Adjustment.} Rather than presenting the AI prediction immediately after the user's initial judgment, task designers should: (1) explicitly show the participant their initial confidence (``You said 'Left' with 6/7 confidence'') before presenting the AI output; (2) present the AI confidence score, not just the label; (3) ask the participant to predict what the AI will say before revealing it (improves metacognitive calibration); (4) allow participants to update their confidence independently before and after seeing the AI explanation, making the confidence shift visible and reviewable.

\textit{Flagging Ambiguous Cases.} At the task level, implement rules: if AI confidence $<$ 0.65 OR if confidence differs dramatically from human agreement levels (e.g., AI high-confidence but annotators only 60\% agree on the label), flag that item for human review or surface disagreement to the user. Communicate explicitly: ``This article received mixed interpretations from domain experts (60\% Center, 40\% Left). The AI predicts Left with 0.42 confidence. Consider whether borderline cases warrant different treatment.'' This prevents false anchoring on uncertain predictions.

\textit{Presentation Order and Cognitive Framing.} Avoid presenting confidence alongside the classification in a way that suggests false certainty. Instead: (1) Present confidence thresholds upfront (``We flag predictions below 0.65 for human review''); (2) show aggregate human agreement rates for similar articles to contextualize the AI confidence; (3) when explanation quality is lower (fewer evidence points, higher perplexity), explicitly note this alongside the classification, reducing false reliance on weak explanations.

\subsection{Ethical Considerations}
Since LLMs are pre-trained on large datasets, researchers have raised ethical concerns about bias and harmful content \citep{Xi2023SafetyAE, Pistilli2022WhatLB}. Annotations and explanations generated by LLMs may reflect underlying biases rather than purely neutral interpretations since LLMs can embed and propagate biases learned during training. In addition, the variability in the quality of AI-generated explanations may have hindered participants' ability to make fully informed decisions, particularly when the explanations were unclear or lacked sufficient detail. In our study, we used only publicly available news articles and relied on OpenAI’s assurance that API data is not stored or used for training. All LLM-generated explanations were manually reviewed before being shown to participants, and no harmful content was observed. We collected no personally identifiable information from MTurk participants, and all participants were fairly compensated.

\subsection{Limitations and Future Work}\label{sec:limitations}
 Although we provide valuable insights in subjective annotation, this work has a few limitations. The complexity of political ideology in the news presents inherent challenges, as it is often highly context-specific. 
We rely on publisher-level labels as reference conventions for this analysis. As documented in Section~\ref{sec:validation_reference_labels}, our validation showed 90\% accuracy against publisher labels, slightly lower than the 97\% reported in prior work \cite{baly2020we} on comparable datasets, which is a difference we attribute to our use of news snippets, our distinct recent news collection, and inter-annotator threshold variations. While this accuracy level still supports the use of publisher labels as reference points, it underscores the inherent subjectivity of bias classification and suggests our findings describe alignment with one labeling convention rather than objective accuracy. 
%
We acknowledge that categorizing U.S. news as ``left,'' ``center,'' or ``right'' represents only one technocratic approach that may not align with all audiences' perceptions of media bias. Exploring other political ideologies and selecting news topics based on their level of controversy, rather than covering a general domain, could offer further insights.
Additionally, we recognize that our sample size could be increased by incorporating repeated measurements for participants.
Since our analysis focused primarily on OpenAI GPT models, the results may not extend to other LLMs. 
For measuring LLM-explanation quality, we acknowledged that more user-centered and semantics-sensitive metrics would offer a deeper assessment.

We measured participants' alignment with AI predictions but did not explicitly measure trust in AI as a distinct construct. Trust in AI, defined as confidence that AI can accurately perform the classification task, is conceptually distinct from familiarity with AI (whether participants have used AI before), which we did measure. Participants with similar AI familiarity may differ substantially in their trust in AI-generated information. If our participant sample contained predominantly high-trust individuals, the observed influence of explanations might be specific to trusting populations and less generalizable to more skeptical audiences. Conversely, if trust varied widely across our sample, the consistency of observed effects across conditions is more robust. We recommend that future work explicitly measure trust in AI as a moderator of explanation effects and test whether baseline trust shapes susceptibility to explanation-induced influence.

Our study presented news articles as snippets or first paragraphs (averaging 77 words) rather than full articles. While this design reflects how many readers encounter news in digital contexts, it introduces an important constraint: participants lacked access to the full textual evidence available to readers who voluntarily choose to read complete articles. This limitation may affect the generalizability of findings in several ways. First, participants' confidence levels may differ if they had access to fuller information that could confirm or contradict AI predictions. Second, in real-world reading, readers can discover counterevidence within longer articles; our snippet-based design prevented this. Third, the cognitive demand and information integration patterns differ between reading brief summaries and full articles. The observed increase in reliance on AI explanations might be attenuated if participants had more opportunity to integrate contradictory information or verify claims independently. We recommend that future research replicate these findings with full-article presentations to assess whether the dual-effect findings (increased confidence and reliance alongside interpretive support) persist in higher-information environments.

Although the switching effect was robust across model specifications, it rests on 51 decision changes, so its magnitude is estimated with wide confidence intervals. In addition, our alignment analyses use publisher-level labels rather than article-level consensus labels. Future studies with more participants and evaluations per article could estimate these effects more precisely and could model annotator disagreement directly.

Despite these limitations, our study opens several avenues for future research for better AI integration across various applications.
Our study presents AI assistance sequentially, with the label and explanation shown together. Because Brief and Detailed outputs were generated separately, the GPT label could also differ between conditions for the same article. Future research could hold the AI label constant across explanation formats and include control conditions that separate the AI label, explanation, and second reading effects to identify which elements drive changes in participants’ decisions. Additionally, future research could explore various forms of news content, such as visual or multimedia elements, and examine whether findings generalize to full-article interpretation rather than snippets. Research in higher-stakes subjective domains (medical diagnosis, legal judgment) could clarify whether dual-effect risks scale with the importance of the decision. Overall, these investigations could advance the responsible design of AI systems in subjective domains and improve user calibration rather than mere reliance.


\section{Conclusion}
We investigated LLMs’ ability to classify news political ideology and examined how AI-generated explanations influence participants’ subjective interpretations. Our results show that GPT-4 achieves competitive accuracy compared to leading supervised models. In a between-subjects study with 124 participants, providing AI explanations increased participants’ self-reported confidence level and often guided them to align with AI-assisted reasoning, and was associated with greater consistency across participants' decisions. However, participants also shifted toward GPT predictions that diverged from publisher-level reference labels, suggesting a possible risk of over-reliance. By documenting both benefits and limitations, our study reveals how different explanations framing interacts with model biases to shape interpretive reasoning. These findings provide practical guidance for designing AI systems that support nuanced judgment while mitigating the amplification of bias or misinformation. To encourage further research, our dataset is publicly available at \url{https://doi.org/10.5281/zenodo.22865164}.

\section*{Acknowledgments}
Contributions from Avinash Chouhan and Joy Mwaria were partially supported by the University of Delaware Undergraduate Research Program. The authors disclose using OpenAI GPT models to benchmark and prepare dataset materials for the human study; all generated content was reviewed and edited by the authors.


\bibliographystyle{ACM-Reference-Format}
\bibliography{reference}

\appendix
\section{Mixed-Effects Model Details}
\label{app:mixed}
Table~\ref{tab:switch_coef} reports the complete fixed-effect coefficients for the decision-switching model. The model with crossed participant and article random intercepts yields identical fixed-effect estimates, because the article variance is estimated at zero (a boundary, or singular, fit); a likelihood-ratio test shows no improvement from adding the article random intercept ($\chi^2(1) = 0.00$, $p > .99$). The participant-only models converged without warnings. The same pattern (article variance of zero; unchanged condition estimates) held for final agreement with the AI, movement toward the AI, and confidence change. Political leaning categories ``Other'' ($n = 3$) and ``Prefer not to say'' ($n = 2$) were combined because of their small size.

\begin{table}[!htb]
\small
\caption{Complete fixed-effect coefficients for the mixed-effects logistic regression predicting decision switching ($n = 372$ decisions, 124 participants, 60 articles; participant random-intercept variance $\sigma^2_{P} = 0.141$; article random-intercept variance $\sigma^2_{A} = 0.000$). Reference categories: Brief condition; Democrat; regular news reader. Initial confidence and AI familiarity are mean-centered. Wald 95\% confidence intervals.}
\label{tab:switch_coef}
\centering
\begin{tabular}{lcccc}
\toprule
\textbf{Term} & \textbf{$b$} & \textbf{SE} & \textbf{95\% CI} & \textbf{$p$} \\
\midrule
Intercept & $-1.295$ & 0.411 & [$-2.101$, $-0.490$] & .002 \\
Detailed condition & 1.172 & 0.429 & [0.331, 2.014] & .006 \\
Initial confidence & $-0.173$ & 0.135 & [$-0.437$, 0.092] & .200 \\
Initial human--AI agreement & $-3.728$ & 0.686 & [$-5.073$, $-2.382$] & $<.001$ \\
AI--reference-label alignment & $-0.137$ & 0.386 & [$-0.893$, 0.619] & .722 \\
Political leaning: Republican & $-0.002$ & 0.511 & [$-1.005$, 1.000] & .996 \\
Political leaning: Independent & 0.006 & 0.474 & [$-0.923$, 0.935] & .990 \\
Political leaning: Other / not stated & $-1.462$ & 1.183 & [$-3.781$, 0.857] & .217 \\
Not a regular news reader & $-0.395$ & 0.543 & [$-1.460$, 0.670] & .467 \\
Prior AI familiarity & 0.243 & 0.145 & [$-0.042$, 0.527] & .094 \\
\bottomrule
\end{tabular}
\end{table}

\end{document}